\documentclass[12pt,preprint]{aastex}
\usepackage{emulateapj5}
\usepackage{xspace}
\usepackage{color}
\newcommand{\chandra}{{\it Chandra}\xspace} 
\newcommand{\xmm}{{\it XMM-Newton}\xspace}

\newcommand{\pulsar}{1E1207\xspace}

\newcommand\lax{\>\vcenter{\hbox{$<$\hskip-.75em\lower1.0ex\hbox{$\sim$}}}\>}
\newcommand\uax{\>\vcenter{\hbox{$>$\hskip-.75em\lower1.0ex\hbox{$\sim$}}}\>}

\slugcomment{Submitted to ApJ} 

\begin{document}

\title{Detailed atmosphere modelling for 
the neutron star 1E1207.4-5209: Evidence of Oxygen/Neon atmosphere}  
\author{Kaya Mori\altaffilmark{1} and Charles J. Hailey\altaffilmark{2}}
\altaffiltext{1}{Canadian Institute for Theoretical Astrophysics, 60 St. George
St., Toronto, ON, Canada, M5S 3H8}
\altaffiltext{2}{Columbia Astrophysics Laboratory, 550 W. 120th St., New York, NY 10027}
\email{kaya@cita.utoronto.ca, chuckh@astro.columbia.edu}
 
\begin{abstract}

We present a comprehensive investigation of the two broad absorption
features observed in the X-ray spectrum of the neutron star
1E1207.4-5209 based on a recent analysis of the 260 ksec XMM-Newton
data by Mori et al. 2005.  Expanding on our earlier work (Hailey \&
Mori 2002) we have examined all previously proposed atmospheric models
for 1E1207.4-5209. Using our atomic code, which rapidly solves
Schr{\"o}dinger's equation for arbitrary ion in strong magnetic field  
(Mori \& Hailey 2002), we have systematically constructed atmospheric
models by calculating polarization-dependent LTE opacities and
addressed all the physics relevant to strongly-magnetized plasmas. We
have been able to rule out virtually all atmospheric models because
they either do not sustain an ionization balance consistent with the
claimed atmosphere composition or because they predict line strengths
and line widths which are inconsistent with the data. Only Oxygen 
or Neon atmospheres at $B\sim10^{12}$ G provide self-consistent
atmospheric solutions of appropriate ionization balance and with line
widths, strengths and energies consistent with the observations.  The
observed features are likely composed of several bound-bound
transition lines from highly-ionized oxygen/neon and they are
broadened primarily by motional Stark effects and magnetic field
variation over the line-emitting region. Further considerations of
plausible mechanisms for the formation of a mid-Z atmosphere likely
rule out Neon atmospheres, and have important implications for the
fallback mechanism in supernova ejecta. Future high resolution
spectroscopy missions such as Constellation-X will be able to resolve
predicted substructure in the absorption features and will measure
magnetic field strength and gravitational redshift independently to
better than 10\% accuracy. 

\end{abstract}


%===========================================================================

\section{Introduction \label{sec_intro}}	

Little is known about the surface composition of isolated neutron
stars (INS). Strong gravity squeezes NS atmospheres to a thin layer
likely composed of a single element \citep{romani87}. Gravitational
stratification forces the lightest element to the top of the
atmosphere on a time scale of 1--100 sec \citep{alcock80}. It is then
likely that the surface is covered by hydrogen. However, various
mechanisms can alter the composition from hydrogen.  During the
supernova explosion, the NS mass cut occurs in the Iron layer
\citep{arnett96}. The surface will be covered by iron if there has
been no fallback nor accretion. Fallback of supernova ejecta or
accretion can put lighter elements on the surface
\citep{woosley95}. The surface composition can evolve with time via
diffuse nuclear burning or excavation by pulsar winds
\citep{chang04_2}. It is impossible to theoretically predict the
surface composition of INS as only a tiny amount of material ($\sim
10^{-19}M_\odot$) is required to constitute the photosphere.

X-ray spectroscopy is the only tool to probe the surface composition
as the surface temperature of observable INS is in the range of $\sim
0.1$--$1$ keV. In particular, detection and identification of spectral
features will uniquely determine the surface composition.
Nevertheless, previous observations of INS by \chandra and \xmm have
somewhat puzzlingly shown no spectral feature \citep{pavlov02_2,
becker02_2}. Only recently, a single absorption feature has been
detected from several nearby radio-quiet NS \citep{haberl03,
vankerkwijk04, haberl04}. Although interpretation of these single
features is still in debate, they probably indicate presence of light
element atmospheres likely composed of hydrogen \citep{vankerkwijk04}.

A point source, 1E1207.4-5209, was discovered near the center of the
supernova remnant PKS 1209-51 by the {\it Einstein} observatory
\citep{helfand84}. The X-ray spectrum is thermal with a temperature
$\sim10^6$ K \citep{vasisht97, zavlin98_2} but no counterpart has been
found at other wavelengths yet \citep{mereghetti96}. Recent detection
of a pulsation period ($P=0.424$ s) provides compelling evidence that
\pulsar is a NS \citep{zavlin00}.

\pulsar is unique among thermally-emitting INS because it
unambiguously shows two absorption features \citep{sanwal02}. It is
the first evidence of a non-hydrogenic NS atmosphere \citep{sanwal02}
and provides a unique opportunity to use spectroscopy to constrain NS
parameters such as magnetic field strength and gravitational
redshift. \citet{mereghetti02} confirmed the two features in the
\xmm/EPIC spectrum. Most recently, \citet{bignami03} and
\citet{deluca04} reported two additional absorption features at 2.1 and
2.8 keV based on the 260 ksec \xmm data. They interpreted the four
absorption features as harmonics of electron cyclotron line at
$B\sim8\times10^{10}$ G. However, \citet{mori05} performed a detailed
analysis on the 2.1 and 2.8 keV feature and found that they are
statistically insignificant and could be related to the instrumental
residuals due to Au M-edges.

Since the discovery, various models have been proposed for the \pulsar
absorption features spanning over a wide range of magnetic field
strengths and surface compositions. They are Helium atmosphere at
$2\times10^{14}$ G \citep{sanwal02, pavlov05}, Iron atmosphere at
$\sim10^{12}$ G \citep{mereghetti02}, mid-Z element atmosphere at
$\sim10^{11}$--$10^{12}$ G \citep{hailey02}, electron cyclotron lines
at $\sim8\times10^{10}$ G \citep{bignami03, deluca04, xu05} and
Hydrogen molecular ions at $(2-6)\times10^{14}$ G \citep{turbiner04}.

In our previous paper \citep{hailey02}, we constrained NS atmosphere
parameter space and found several solutions that are consistent with
the \chandra data. All the solutions were attributed to mid-Z
elements. In the present paper we perform a more detailed theoretical
analysis of NS atmosphere based on our spectral analysis of the 260
ksec \xmm data \citep{mori05}. We concluded that the only plausible
solution is an Oxygen or Neon atmosphere at $B\sim10^{12}$ G.

The outline of the paper is as follows. In \S \ref{sec_obs} we briefly
review our spectral analysis of the 260 ksec \xmm data and illustrate
the key observables for identifying the spectral features. In \S
\ref{sec_atmos} we show that modelling of the non-hydrogenic
atmosphere using our new atomic code for high magnetic field regime is
necessary to interpret the spectral features, in contrast to the
previous efforts.  We also define NS atmospheric parameter space for
our analysis and justify the LTE assumption. In \S\ref{sec_pre} we
review atomic physics in the strong magnetic field regime, focusing on
feasible spectral features in the X-ray band.

Three major steps follow to constrain the NS parameter space defined
in \S\ref{sec_atmos}. Firstly, we apply the atomic physics methodology
to find possible solutions that are consistent with the observed
feature location (\S\ref{sec_method}). Some of the solutions were
presented in our previous paper \citep{hailey02}. Secondly, among the
potential solutions obtained in \S\ref{sec_method}, we construct an
Oxygen atmosphere model at $B=10^{12}$ G (\S\ref{sec_oxygen}). As we
will show, this is the only solution consistent with the observed line
parameters. Following the partially-ionized magnetized Hydrogen
atmosphere models \citep{potekhin03}, we calculate LTE opacities by
taking into account all the physics required for the NS atmosphere
problem; atomic physics in strong B-field, ionization balance, line
broadening and polarization vectors.  Thirdly, we critically analyze
the other proposed models as well as the other possible solutions from
\S\ref{sec_method} (\S\ref{sec_others}). All the models except O/Ne
atmosphere at $B\sim10^{12}$ G are ruled out based on the NS
atmosphere physics addressed in \S\ref{sec_oxygen} or the existing
spectral models with full radiative transfer solutions. Finally, we
discuss implications of O/Ne atmosphere in \S\ref{sec_imply} and
summarize our results in \S \ref{sec_last}.

%=============================================================================

\section{Data analysis of the 260 ksec \xmm observation  \label{sec_obs}} 

In this section, we briefly summarize our results from the analysis of
the 260 ksec \xmm/EPIC data. More details can be found in
\citet{mori05}.

In \citet{mori05}, we found that two continuum components ($C1$ and
$C2$) and two absorption lines ($L1$ and $L2$) are required to fit the
data. $C1$/$L1$ and $C2$/$L2$ refer to the lower and higher energy
spectral component respectively.  The three classes of models are
considered (with the notation $C*L$ meaning line $L$ resides on
continuum component $C$)
\begin{eqnarray} 
\mbox{Model I:} && C1*L1*L2+C2\nonumber\\ \mbox{Model II:} &&
C1*L1+C2*L2 \nonumber \\ \mbox{Model III:} && (C1+C2)*L1*L2 \nonumber
\end{eqnarray}
Model I assumes that $L1$ and $L2$ are from the same region as the
emission of $C1$ and Model II assumes that they are from different
regions. Model III assumes that $L1$ and $L2$ are from a layer above
the two regions emitting continuum photons.

\begin{figure}
\epsscale{1.0} 
\plottwo{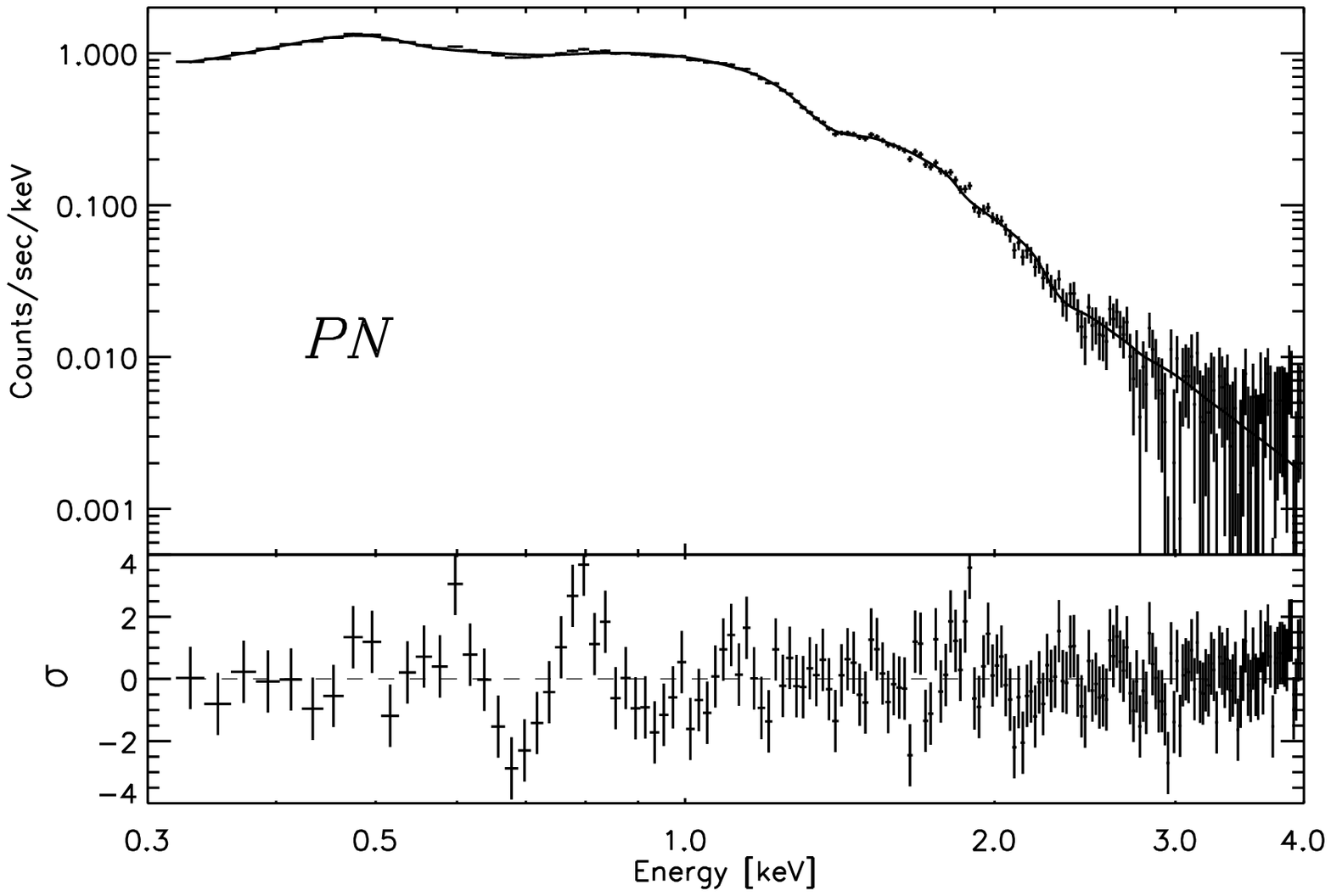}{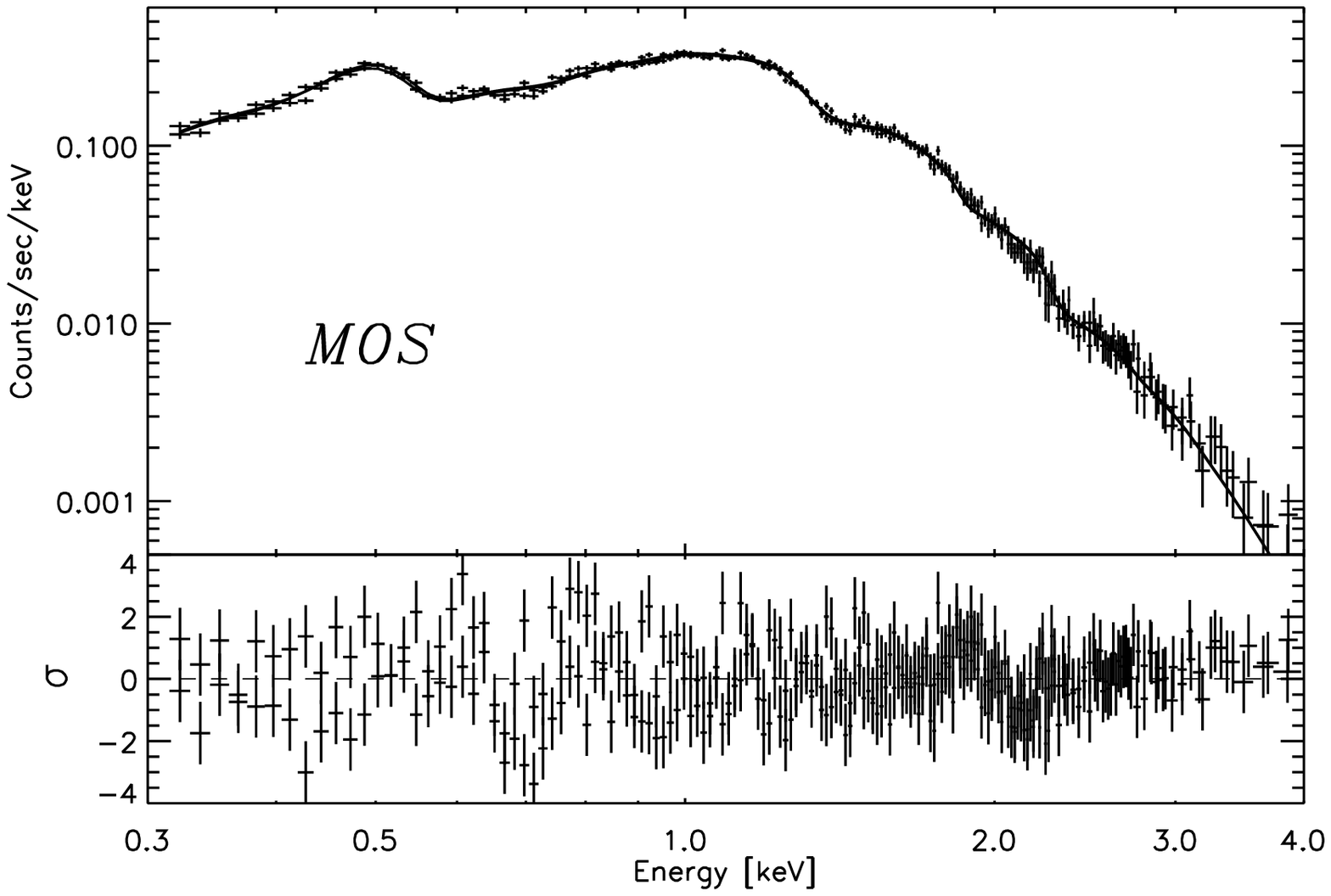}
\caption{\xmm/EPIC PN-singles (left) and MOS (right) spectrum and
residuals. We fit the spectra with model I with two blackbody
components and two Gaussian absorption lines at $\sim0.7$ and
$\sim1.4$ keV. For the MOS spectrum, we combined MOS1 and MOS2 data to
improve photon statistics. An interesting substructure is seen around
0.7 keV in both PN and MOS spectrum. \label{fig_data}}
\end{figure}

All the two-component thermal models composed of either blackbody and
magnetized Hydrogen atmosphere model fit the data well, yielding a
chi-squared very close to unity. For the fit parameters, the reader
can refer to table 1 and 2 in \citet{mori05}. Figure \ref{fig_data}
shows the PN singles spectrum fitted by model I with two blackbody
components and two Gaussian absorption lines at 0.7 and 1.4 keV. The
residuals appear more or less the same for the different fit models
and all show an interesting substructure around 0.7 keV in both the PN
and MOS data \citep{mori05}.  Continuum models with power-law
components are ruled out because they do not adequately fit the data,
leaving significant residuals above 2 keV. When we fit
photo-absorption edges to the two absorption features at 0.7 and 1.4
keV, the chi-squared was significantly larger. Therefore we rule out
models with photo-absorption edges.

Among the three models described above, we adopt model I to
investigate the two absorption features. Model I describes a spectrum
composed of a thermal component from a large area on the surface with
two absorption features and an additional hot thermal component from a
smaller area. The fitted temperature is $\sim150$--200 eV for the cold
thermal component (Note that we converted the best-fit temperature to
the NS frame by a gravitational redshift factor $(1+z)=1.3$ which
corresponds to $M=1.4M_\odot$ and $R=10$ km). The errors in the fitted
temperature come primarily from different continuum model fits to the
data either by blackbody or magnetized Hydrogen atmosphere models. We
found model II implausible from our phase-resolved spectroscopy
\citep{mori06_1}. Theoretical models for model III can be found
elsewhere \citep{dermer91, wang98, ruderman03} and we will discuss
several cases associated with model III later.

Hereafter we list important observables for identifying the absorption
features.

\begin{itemize} 

\item[(a)] Equivalent widths of the two features are comparable.

\item[(b)] Line energy ratio of the two features is $\sim1.9$

\item[(c)] Line energies of the two features are $\sim0.75$ keV and
$\sim1.4$ keV.

\item[(d)] Line widths ($\equiv\Delta E /E$) are $\sim$40\% and
$\sim$15\% for the 0.7 and 1.4 keV feature.

\end{itemize} 

%%%%%%%%%%%%%%%%%%%%%%%%%%%%%%%%%%%%%%%%%%%%%%%%%%%%%%%%%%%%%%%%%%%%%%%%%%%%%

\section{Magnetized non-hydrogenic atmosphere modelling \label{sec_atmos}}

In principle, the thermal spectrum from a NS is {\it never} a
blackbody.  The atmosphere modifies the spectral energy distribution
(SED) and results in the formation of features in the photospheric
spectrum. In general, NS atmosphere models are constructed in the
parameter space of magnetic field strength ($B$), surface element
($Z$), effective temperature ($T_{eff}$) and gravitational
acceleration ($g$) \citep{becker02}.

\subsection{Previous approaches: inapplicable to \pulsar}  

In some special cases, self-consistent NS atmosphere models with full
radiative transfer solutions have been constructed with sufficient
accuracy for fitting the current X-ray spectroscopic data.  However,
as we will show, these approaches are not applicable to \pulsar and
magnetic non-hydrogenic atmosphere models are required to interpret
the observed spectral features.

Non-magnetic atmosphere models assuming $B=0$ are applicable to NS
with $B<10^9$ G, such as milli-second pulsars \citep{zavlin02_2} and
accreting NS \citep{rutledge99}. At $B<10^9$ G, atomic data for $B=0$
case are sufficiently accurate since magnetic field effects do not
perturb atomic structure significantly (more details in \S
\ref{sec_pre}).  Sophisticated non-magnetic atmosphere models have
been constructed for various atmosphere compositions by implementing
reliable opacity libraries such as OPAL
\citep{zavlin96,rajagopal96,gansicke02}. For any surface composition,
non-magnetic atmosphere models do not produce spectral features at 0.7
and 1.4 keV and this is spectroscopic evidence that \pulsar possesses
a magnetic field in excess of $10^{9}$ G.

At $B>10^9$ G, modeling NS atmospheres is significantly more difficult
because of the complicated magnetic effects on atomic structure and
radiation (refer to \citet{lai01} for review). In the high B-field
regime, only Hydrogen atmospheres have been studied in great detail
\citep{pavlov95_1, potekhin99, potekhin03,
potekhin04}. Partially-ionized Hydrogen atmosphere models show
spectral features in the soft X-ray band such as proton cyclotron
line, photo-ionization edge or atomic transition lines at
$B\uax10^{13}$ G \citep{ho03, ho04}. However, \pulsar shows absorption
features that Hydrogen atmosphere models cannot reproduce. More
specifically, the 1.4 keV absorption feature cannot be produced by
Hydrogen atoms as the binding energy of a Hydrogen atom never exceeds
$\sim1$ keV at any B-field \citep{sanwal02}.

Therefore, a non-hydrogenic magnetized atmosphere model is required to
fit the spectral features present in \pulsar \citep{sanwal02}
\footnote{More precisely, the spectral features of \pulsar cannot be
explained by Hydrogen atom. If Hydrogen molecular ions such as H$_2^+$
and H$_3^{2+}$ exist in the atmosphere, they can produce spectral
features at the observed energies when $B>10^{14}$ G
\citep{turbiner04}. }. However, the few existing non-hydrogenic
atmosphere models \citep{miller92, rajagopal97} are far from complete
mainly due to a lack of accurate atomic data for multi-electron ions
in the high magnetic field regime.

The model of \citet{miller92} and \citet{rajagopal97} which utilizes a
one-dimensional Hartree-Fock method \citep{neuhauser87} (hereafter
1DHF) is rather crude and the energy values and oscillator strengths
have as much as 10\% and a factor of 2 uncertainties,
respectively. The 10\% accuracy of the 1DHF code in transition
energies can possibly constrain surface element, but determination of
$B$ and $z$ will be inaccurate since they are sensitive to errors in
the computed and measured transition energies. The accuracy is
unacceptable at the magnetic field strengths and mid-Z elements of
interest to us since the 1DHF neglects the effects of excited Landau
levels.

Similar problems plague other atomic codes developed for the strong
magnetic field regime \citep{jones_md99_1,ivanov00}. They are highly
accurate, providing binding energies to better than 0.1\% , but their
computational speeds are extremely slow because they entail solving
the 2-dimensional Schr{\"o}dinger equation (for detailed comparison of
these different atomic calculations, refer to \S 4 in
\citet{mori02}). They calculate ground state energies of low-Z atoms,
but do not provide oscillator strengths. However assessing NS
atmospheric conditions requires both transition energies and
oscillator strengths for various binding states over a large set of NS
parameters.

\subsection{Our approach: opacity calculation based on accurate atomic data} 

Here we utilize an approach for identifying the spectral features
based on accurate and extensive atomic data composed of transition
energies and oscillator strengths. Our atomic code suitable for the
high magnetic field regime provides better than 1\% accuracy in
transition energies (more details can be found in \citet{mori02} and
\S\ref{sec_pre}).

In NS atmosphere spectroscopy, identification of spectral features is
further complicated by gravitational redshift ($z$) as it lowers line
energies in the observer's frame by a factor of $(1+z)$. Therefore,
spectral feature location depends on the four parameters; $B$, $Z$, 
number of bound electrons ($n_e$) and
$z$. Spectral features in the X-ray band emerging from
strongly-magnetized plasmas are either cyclotron lines, atomic
transition lines or photo-absorption edges. In \S\ref{sec_method} we
investigated any possible combination in the $(B,Z,n_e,z)$ phase space
which generates the observed location of the features. As a result,
seven solutions in the $(B,Z,n_e,z)$ phase space were found by the
atomic physics methodology. Derivation of more than one solution is
not surprising because we attempt to match two feature locations in
the 4-dimensional parameter space of $(B,Z,n_e,z)$.

One can constrain the NS parameter space further in light of
self-consistency and physical plausibility of the models. A
self-consistent atmosphere model must solve radiative transfer
equations with reliable opacity tables and equation of state as was
done for magnetic Hydrogen atmosphere models. Line identification
based on transition energies and oscillator strengths presented in
\S\ref{sec_method} is inadequate for NS atmosphere spectroscopy.  In
the present work, we calculate LTE opacities by taking into account
ionization balance and dielectric properties in strongly-magnetized
dense plasmas. In the case of \pulsar, opacity highly constrains the
NS atmosphere parameter space and leads to a single solution that is
physically plausible and consistent with the observed line parameters.
As a next step, radiative transfer is in progress using the LTE
opacities to construct self-consistent spectral models for fitting to
the \xmm data \citep{mori06_2}. Nevertheless, our conclusion on the
surface composition is unchanged regardless of radiative transfer
since our solution is consistent with the observed line parameters
over a large range of the NS atmosphere parameters.
  
Hereafter we briefly review each process involved in computing LTE
opacities.  In a strong magnetic field, radiation becomes highly
anisotropic and propagates in two normal polarization modes
\citep{ginzburg70, meszaros92}. Therefore, opacities are dependent on
the direction of photon propagation and the polarization mode. Under
the LTE assumption, the absorption opacity $\kappa^j (E, \theta_B,
\rho, T)$ [cm$^2$/g] for a photon with energy $E$ propagating in a
polarization mode ($j=1,2$) and at an angle $\theta_B$ 
relative to magnetic field is written as \citep{potekhin03},
\begin{equation} 
\kappa^j (E, \theta_B, \rho, T) = M^{-1} \sum_\alpha \sum_i
|\vec{e}_\alpha^j (E, \theta_B)|^2 \sigma^i_\alpha(E) f_i(\rho, T).
\end{equation}
$M$ is the atomic mass. $\sigma^i_\alpha (E)$ is the cross section for
an ionization state $i$ and a basic polarization mode $\alpha$. The
three basic (cyclic) polarizations ($\alpha=0, \pm1$) are defined with
respect to the magnetic field (one linear along B-field and two
circular transverse to B-field). $\sigma^i_\alpha (E)$ is a quantity
derived directly from the Schr{\"o}dinger equation using our atomic
code. $f_i$ is the fraction of ionization state $i$, which depends
only on plasma density ($\rho$) and temperature ($T$) under the LTE
assumption. $f_i$ is calculated by solving the Saha-Boltzmann
equations modified for strong magnetic regime. $|\vec{e}_\alpha^j|$ is
the projection of the basic polarization vectors in the normal modes,
reflecting the dielectric properties of plasma in the magnetic
field. We calculate the polarization vectors using the Kramers-Kronig
relations adopted for partially-ionized Hydrogen atmosphere models
\citep{bulik96, potekhin04_2}. For some cases, information on the line
widths helps to check the plausibility of the models. Various line
broadening mechanisms are investigated quantitatively. We emphasize
that the procedures presented in this paper were justified and used
for constructing partially-ionized Hydrogen atmosphere models in
strong magnetic field regime \citep{potekhin03, potekhin04}. All the
physics relevant to strongly-magnetized dense plasmas are addressed in
our opacity calculations following \citet{potekhin03}.

%%%%%%%%%%%%%%%%%%%%%%%%%%%%%%%%%%%%%%%%%%%%%%%%%%%%%%%%%%%%%%%%%%%%%%%%%%%%%%

\subsection{Range of the atmospheric parameters considered in our analysis \label{sec_param_space}}

We define the range of NS atmospheric parameters considered in our
analysis.

\subsubsection{Magnetic field strength \label{sec_bfield}}

We consider a single value of B-field to identify the observed
spectral features. This assumption is valid within a thin photosphere
on the NS surface.  However, magnetic field can vary over the
line-emitting area thus producing line broadening. We will discuss
line broadening due to non-uniform B-field distribution in
\S\ref{sec_nonuniform}. We investigate a range of magnetic field
strength from $B=10^{11}$--$10^{15}$ G in the following analysis. This
is the range where our atomic code accurately predicts spectral
features for elements $Z=1$--$26$ in the X-ray band. There are several
cases which require B-field smaller than $10^{11}$ G, but we can rule
out these cases from simple atomic physics arguments.

\subsubsection{Atmospheric composition \label{sec_composition}}

We consider atmospheres composed of both pure elements and mixtures of
different elements in the range of $Z=1$--$26$. We exclude rare
elements expected in the products of explosive nucleosynthesis in
supernova explosions (e.g. for abundant elements in supernova ejecta,
refer to \citet{thielemann96}).

\subsubsection{Gravitational redshift \label{sec_redshift}}

We set the range of possible gravitational redshift to
$z=0$--$0.85$. The upper bound originates in constraints associated
with the nuclear EOS including stability ($dP/d\rho>0$) and causality
($dP/d\rho<c^2$ where $c$ is the speed of light) conditions inside the
star \citep{lindblom84, haensel99}.

\subsubsection{Photosphere temperature \label{sec_temperature}}

Effective temperature measured from overall spectral fitting is
sensitive to assumed atmosphere composition. For \pulsar,
$kT_{eff}=150$-$200$ eV covers the whole range of effective
temperature.  The effective temperature merely represents the total
thermal flux, but in reality a NS atmosphere has a temperature
gradient under hydrostatic equilibrium. Recent atmosphere modelling
showed that absorption features are formed at shallow depths in the
photosphere ($\sim(10^{-3}-10^{-1})\tau_{T}$ where $\tau_{T}$ is the
Thompson depth). Temperatures at the line forming depths are lower
than the effective temperature at most by a factor of few \citep{ho01,
miller92}. Therefore, we consider a temperature range of $\sim50$--200
eV. Later in our opacity calculations, we adopt $kT=150$ eV because 
mid-Z atmospheres generally have a temperature gradient close to the
grey profile as bound-bound and bound-free opacities are more 
important than free-free opacities \citep{miller92, rajagopal97}. 

\subsubsection{Plasma density \label{sec_rho}}

NS photosphere density is roughly in the range of $\sim10^0$--$10^2$
 g/cm$^{3}$ depending on surface composition, magnetic field and
 polarization mode \citep{pavlov95_1, becker02, miller92,
 rajagopal97}. Since absorption lines are formed at smaller optical
 depths in the photosphere, we focus on a much smaller plasma density
 range. Hereafter, we consider a density range of
 $\rho\sim10^{-4}-10^{2}$ g/cm$^{3}$ based on \citet{ho01},
 \citet{ho03} and \citet{miller92}.

\subsubsection{LTE consideration \label{sec_lte}}

Before proceeding to identify the observed features, we investigate
the validity of the local thermodynamics equilibrium (LTE)
assumption. LTE greatly simplifies calculation of ionization states
and level population as they depend only on plasma density and
temperature.  In a high density plasma where collisional processes are
faster than radiative processes, LTE is well established even for the
innermost electron.  LTE is achieved when there is simultaneously (1)
a Maxwellian electron distribution (2) Saha equilibrium (3) Boltzmann
equilibrium \citep{salzmann98}.  The conditions for a Maxwellian
electron distribution are fulfilled at quite low density since the
electron self-collision time is very short \citep{spitzer62}.  The
range of baryon density considered in our analysis well satisfies
conditions for the Saha and Boltzmann equilibrium at $kT=50$--200 eV
\citep{griem64, salzmann98}, hence the NS atmosphere is likely in LTE.

%%%%%%%%%%%%%%%%%%%%%%%%%%%%%%%%%%%%%%%%%%%%%%%%%%%%%%%%%%%%%%%%%%%%%%%%%%%%%%

\section{Preliminaries for NS atmosphere spectroscopy: Atomic physics in the Landau regime \label{sec_pre}}

Before we proceed to constrain the NS atmosphere parameters, we review
important magnetic field effects directly relevant to location,
strength and width of the observed features.

In a strong magnetic field, atomic structure is quite different from
the $B=0$ case. A strong magnetic field deforms the atom to a
cylindrical shape when magnetic field effects are larger than Coulomb
field effects. This regime (often called the Landau regime) is defined
as $\beta_Z>1$ where $\beta_Z=B /4.7\times10^{9}Z^2$. In the Landau
regime the binding energy of bound electrons increases
significantly. For instance, the ionization threshold of a Hydrogen
atom is 160 eV at $B=10^{12}$ G.

A bound electron in the Landau regime is often denoted by two quantum
numbers $m$ and $\nu$ (hereafter we use $(m\nu)$ to denote bound
states). $m$ is a magnetic quantum number and $\nu$ is a longitudinal
quantum number along the field line.  There are two additional quantum
numbers, Landau number ($n$) and electron spin component along the
field ($s$). They are fixed to $n=0$ and $s=-1/2$ in the Landau
regime, where electron cyclotron energy far exceeds Coulomb energy and
thermal energy.  $\nu=0$ states have larger binding energy
(tightly-bound states) and $\nu>0$ states have smaller binding energy
(loosely-bound states). In most cases, the ground state configuration
is composed of bound electrons in tightly-bound states i.e. $(m0)$
states. The innermost electron is in a $(00)$ state and binding energy
of tightly-bound states decreases as $m$ increases. Excited states
often involve loosely-bound states (i.e. $(m\nu)$ states with $\nu>0$)

The effects of finite nuclear mass are mentioned here since they help
determine the location and width of atomic features. In the absence of
a magnetic field, the center-of-mass term and the nucleus-electron
term in the Hamiltonian are separated by a canonical
transformation. When a magnetic field is present, collective motion of
an atom and its internal electric structure are coupled
\citep{pavlov93, potekhin94} and there remain two additional terms in
the transformed Hamiltonian, the nuclear cyclotron term and the
motional Stark term (which will be discussed in \S \ref{sec_ms}).
Essentially both terms lower the binding energy from that computed
from the Hamiltonian with infinite nuclear mass. The nuclear cyclotron
term ($=m\hbar\omega_{NB}$ where $\omega_N=ZeB/m_Nc$, $m_N$ is the
nuclear mass and $c$ is the speed of light) can be simply added to
solutions for the infinite nuclear mass case since the term commutes
with the Hamiltonian. The combination of the two effects auto-ionizes
bound electrons at high magnetic fields \citep{potekhin94,
kopidakis96}. Hereafter electron binding energies are computed with a
correction for the nuclear cyclotron term. Essentially, the correction
to energies (hence oscillator strengths) of $\Delta m \ne 0$
transition lines becomes large at $B>10^{14}$ G, while it is
negligible at $B\sim10^{12}$ G.

\subsection{Feasible atomic features in the X-ray band \label{sec_xray}} 

Feasible atomic transitions in the X-ray band are photo-absorption
edges from tightly-bound states or photo-absorption lines from
tightly-bound states to tightly-bound states (hereafter tight-tight
transition) or loosely-bound states (hereafter tight-loose
transition). Landau transitions (between different $n$) and spin-flip
transitions (between different $s$) do not occur in a strong magnetic
field since they require photon energies comparable to the electron
cyclotron energy ($=\hbar\omega_{eB}=11.6B_{12}$ keV and
$B=10^{12}B_{12}$ G) to excite the ground states. Nevertheless, we
will consider $\Delta n \ne 0$ transitions (the so-called cyclotron
lines) as well.  An electron spin-flip transition with $\Delta s =1$
has essentially the same line energy as a $\Delta n =1$
transition. However, the line strength of spin-flip transition is
significantly smaller than a $\Delta n =1$ transition
\citep{melrose81}. Therefore we do not consider the spin-flip transition.

\begin{figure}
\epsscale{0.7} \plotone{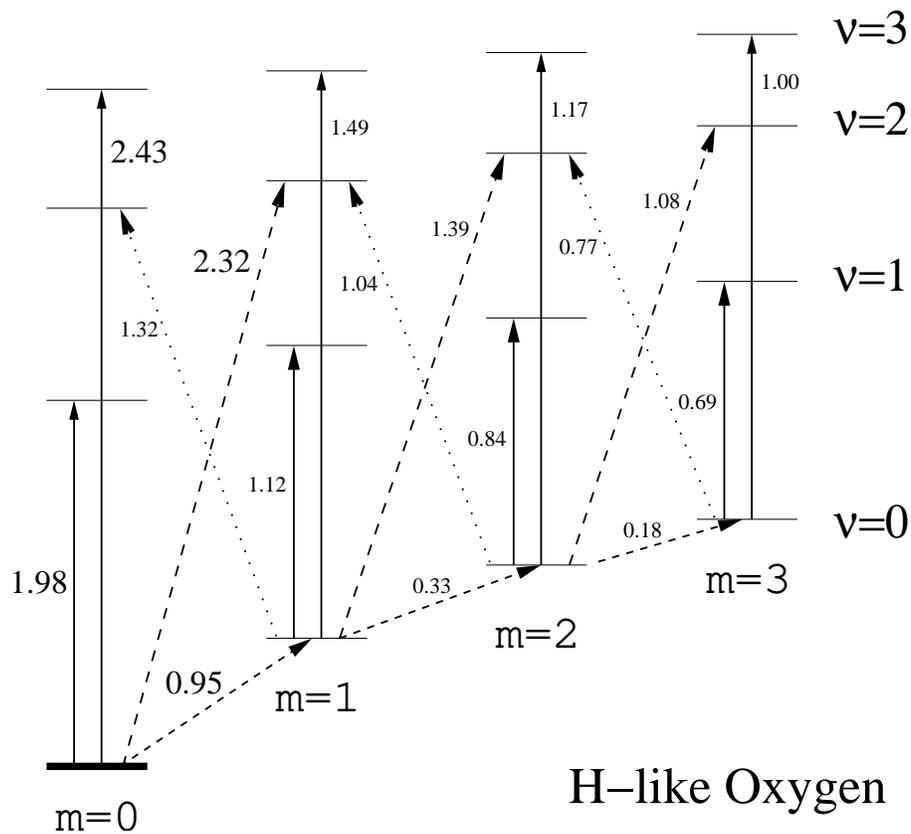}
\caption{Grotrian diagram for strong transitions in the Landau
  regime. The solid, dashed and dotted lines denote transitions for
  $\alpha=0, +1$ and $-1$ mode respectively. (Unredshifted) transition
  energies [keV] of
  H-like oxygen at $B=10^{12}$ G are also shown next to the
  arrows (large and small font size correspond to 
 transition energies from ground states and excited states
 respectively). The ground state ($(00)$ state for H-like oxygen) is
  indicated by a bold horizontal line. 
\label{fig_lines}}
\end{figure}

For bound-bound transitions, $\Delta m = 0$ and $\Delta\nu=odd$ or
$\Delta m=\pm 1$ and $\Delta\nu=even$ transitions are allowed by
dipole selection rules (i.e. they have large oscillator strengths) The
former transitions occur in the longitudinal polarization mode
($\alpha=0$, parallel to B-field), while the latter transitions occur
in the circular polarization mode ($\alpha=\pm 1$, transverse to
B-field). Figure \ref{fig_lines} shows a Grotrian diagram for strong
bound-bound transitions in the Landau regime along with line energies of a
H-like oxygen ion at $B=10^{12}$ G. 

\begin{figure}
\epsscale{1.0} \plotone{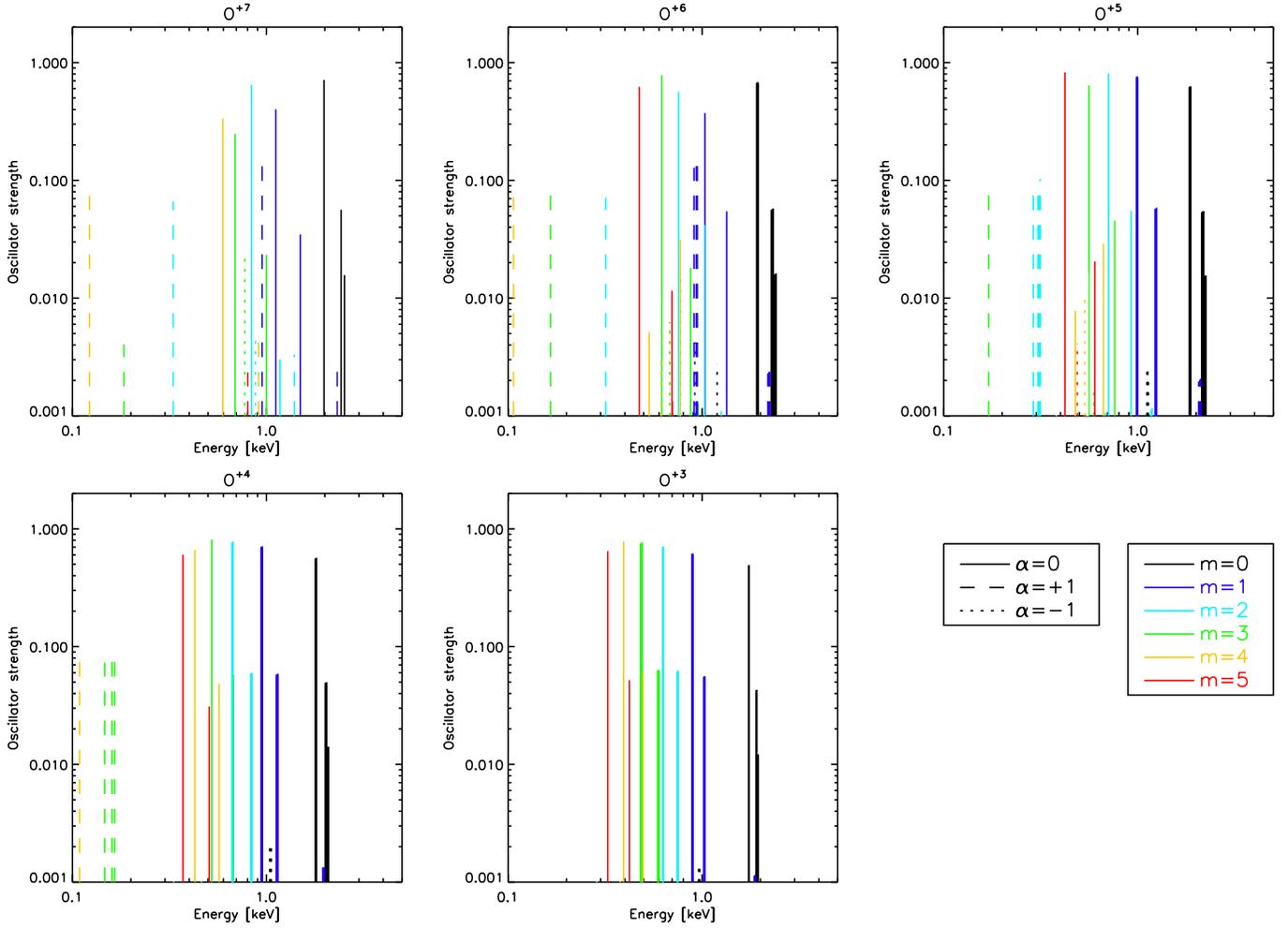}
\caption{Line spectra of five ionization states of  oxygen in the X-ray band at
$B=10^{12}$ G. Note that energies are unredshifted. Similar to figure \ref{fig_lines}, different line types
  indicate transitions for different cyclic polarization modes. The
  lines are denoted with different colors according to initial $m$ of
  a transiting bound electron. \label{fig_os}}
\end{figure}

To illustrate transition lines for different polarization modes and
initial states, line spectra of five
different oxygen ions at $B=10^{12}$ G are shown in figure \ref{fig_os}. Strong magnetic
field separates different $m$ states and they make groups of
transition lines according to different $m$ (denoted by different
colors) as seen in figure \ref{fig_os}. In this sense, transition lines are {\it evenly}
distributed in the energy space. This contrasts with the zero magnetic
field case in which transitions from the K- and L-shell are largely
separated (e.g. 6 keV and 1 keV for K- and L-shell transitions of
Iron). In each $m$ group, several tight-loose transitions are feasible
but the oscillator strength decreases as $\nu$ becomes larger
\citep{ruder}, while in general oscillator strengths for lines depend
weakly on $m$ \citep{miller92, mori02}. Separation between the groups of lines becomes smaller
as $m$ increases. The line distribution shown in figure
\ref{fig_os} is more or less the same for different elements in the Landau regime.  We will apply these
properties for identifying the observed features from line location
and line energy ratio.

\subsection{Our atomic code: Multi-configurational, perturbative, hybrid,
Hartree, Hartree-Fock method \label{sec_code}}

Identification of atomic features is a time-consuming task requiring a
systematic search of the $(B,Z,n_e,z)$ parameter space.  For the
purpose of identifying atomic features, we developed a fast and
accurate atomic code based with an approach we call
multi-configurational, perturbative, hybrid, Hartree, Hartree-Fock
theory (hereafter MCPH$^3$) \citep{mori02}. For most electron
configurations and magnetic field strengths relevant to \pulsar, the
MCPH$^3$ code calculates transition energies and oscillator strengths
for any given electron configuration of an arbitrary ion in the Landau
regime. The MCPH$^3$ code provides atomic data to better than 1\% and
10\% accuracy for energies and oscillator strengths respectively.  The
MCPH$^3$ code achieves significantly faster computation time compared
to other atomic codes, such as multi-configurational Hartree-Fock
codes \citep{ruder} and two-dimensional Hartree-Fock codes
\citep{ivanov00}, by use of its perturbation method. The resultant
fast algorithm of the MCPH$^3$ code enables us to compute atomic data
over a large set of $B, Z$ and $n_e$ with reasonable CPU time. For
instance, computation of a ground state configuration of Oxygen atom
(8 bound electrons) was achieved within 5 min on Celeron with 2.4 GHz
CPU and 512MB memory. The computation time is roughly proportional to
$n_e^2$ where $n_e$ is the number of bound electrons
\citep{mori02}. We checked that convergence of our atomic calculation was 
 achieved to better than 0.1\% by monitoring Hartree energy of each electron
 orbital. We also extended our atomic code to calculate
photo-absorption cross sections and perform molecular structure
calculation in the Born-Oppenheimer approximation. Some of the new
results are used to constrain the atmosphere models for \pulsar.

\subsubsection{Validity range of our atomic code \label{sec_validity}} 

By applying perturbation methods to higher Landau levels the MCPH$^3$
code has an extended range of applicable magnetic field.  The MCPH$^3$
code provides useful results for our analysis (i.e. accuracy in energy
of a few\%) for $\beta_Z\ge0.3$. This was determined by comparing it
with atomic data generated by other approaches
(e.g. multi-configurational Hartree-Fork code by \citet{ruder}) known
to be highly accurate in the intermediate magnetic field regime
($\beta_Z\sim0.1$--$1$).

On the other hand, relativistic effects become significant at high
magnetic field and may affect results from the MCPH$^3$ code, which
solves the non-relativistic Schr{\"o}dinger equation. Transverse
motion of electrons around the magnetic field becomes relativistic at
$B>4.4\times10^{13}$ G where the electron cyclotron energy is equal to
the electron rest mass energy. The wavefunction of electrons in their
transverse direction (the Landau function) has the same form as in the
non-relativistic case, as long as electrons are in the ground state of
the Landau levels. This is true for the high magnetic fields where the
transverse electron motion is relativistic \citep{lai01}. Therefore,
relativistic corrections to the transverse motion are irrelevant. On
the other hand, bound electrons are subject to a larger nuclear
Coulomb field since they are closer to the nucleus at higher magnetic
field. Accordingly, relativistic effects significantly affect the
longitudinal motion of bound electrons.  However, relativistic effects
due to the nuclear Coulomb field are negligible as long as the energy
range of interest (in our case $\sim1$ keV) is much smaller than the
electron rest energy \citep{angelie78}.

As we do not include higher than the 1st order perturbative terms in
 the electron-electron exchange energy, our atomic calculation becomes
 less accurate for electron configurations where bound electrons are
 closer to each other. This is the case for large $m$ states in ground
 state configurations of ions with many electrons. Accuracy in
 transition energies and oscillator strengths can drop $\sim$10\% and
 $\sim$50\%. However, it does not affect determination of NS
 atmosphere parameters because the spectral features are due to
 highly-ionized O/Ne where uncertainty associated with the exchange
 term is tiny ($<0.1$\%).

%==============================================================================

\section{Atomic physics methodology of identifying absorption features \label{sec_method}}

We can constrain the $(B, Z, z, n_e)$ phase space from the observables
 presented in \S \ref{sec_obs} and the relevant physical effects
 discussed in \S \ref{sec_pre}. We briefly outline our three-step
 procedure in connection with the observed properties listed in
 \S\ref{sec_obs}. First, we constrain ionization state ($n_e$) from
 the observation (a), transition energies, oscillator strengths and
 level population under the LTE assumption. Secondly, we present a
 scheme to determine magnetic field strength ($B$) from the
 observation (b). We adopt the line energy ratio as a strong
 diagnostic parameter to determine $B$ for a given $Z$. $z$ is
 separately determined from the other parameters since binding
 energies depend on $B, Z$ and $n_e$ in a non-linear fashion, while
 the gravitational redshift lowers the location of the features
 linearly by a factor of $(1+z)$.  This is different from the case of
 cyclotron lines where $B$ and $(1+z)$ shift spectral features
 identically. Finally, we determine surface element ($Z$) from the
 observation (c), requiring a fit to the features in the \xmm data
 with a reasonable range of $z$.

Following our previous paper \citep{hailey02}, we consider three
cases; pure atomic transitions (Case A), pure cyclotron transitions
(Case B) and a combination of atomic transition and cyclotron line
(Case C). In this section, our discussion is mostly devoted to case
A. For case B and C, we merely present all the possible solutions and
defer more detailed investigation on their plausibility to
\S\ref{sec_others}.

The parameter space where we search for plausible solutions is vast
and 4-dimensional. Several simple and reasonable assumptions help to
significantly reduce the amount of work and time required. Later, we
relax the assumptions and consider how this changes the results. We
assume a single element and a single ionization state as the zero-th
order constraints. The former condition is rather realistic as the
gravitational sedimentation is fast on the NS surface. Later we will
investigate whether admixtures with more than one element are
consistent with the data. In the next section, we relax the latter
condition by allowing more than one ionization state while ionization
balance will be investigated in \S\ref{sec_density}.

\subsection{Case A: pure atomic transition lines \label{sec_caseA}} 

In this section we focus on bound-bound transitions for the following
reasons, although we considered all combinations of spectral features
including bound-free transitions. First, models with photo-absorption
edges do not fit the \xmm/EPIC data \citep{mori05}. Second, bound-free
transitions have significantly smaller cross sections (hence
opacities) than bound-bound transitions in the most plausible case
(i.e. O/Ne atmosphere at $B\sim10^{12}$ G) (\S\ref{sec_cs}).

\subsubsection{Constraining ionization state  from condition (a) \label{sec_n_e}} 

First of all we constrain the ionization state ($n_e$) from the
observation (a) that the two features have similar strengths.  We
examine oscillator strengths of atomic transition lines which may
appear in the X-ray band as well as level population from the
Boltzmann distribution.

\subsubsubsection{H-like ion}

Since the number of observed absorption features is few, we begin with
H-like ions. See figure \ref{fig_lines} for a Grotrian diagram of
H-like oxygen at $B=10^{12}$ G. Ground state configuration of H-like ions consists of
$(00)$ state. There are two strong lines from the ground state:
$(00)\rightarrow(01)$ and $(00)\rightarrow(10)$ transition. The former
is the strongest among tight-loose transitions and its oscillator
strength is close to unity. The latter is $\Delta m =+1$ tight-tight
transition and its oscillator strength decreases with B-field as
$B^{-1}$ \citep{ruder}. In the Landau regime, a line energy ratio of
these two transitions is in the range of $\sim1.5$--3.0 (see
\S\ref{sec_b_const} and figure \ref{fig_ratio}), which covers the
observed line energy ratio $\sim1.9$.

Other tight-loose transitions such as $(00)\rightarrow(0\nu)$ with
$\nu = 3, 5 ...$ or $(00)\rightarrow(1\nu)$ with $\nu = 2, 4 ...$ have
line energy overlapping with the 0.7 and 1.4 keV feature. However,
their oscillator strengths decrease as $(\nu+1)^{-3}$ \citep{ruder}
and high $\nu$ states are possibly destroyed by pressure ionization
(\S\ref{sec_density}). Therefore tight-loose transitions to higher
$\nu$ states are weaker than $\Delta\nu=1$ transitions. This is also
true for other ionization states.

The 1st excited state of a H-like ion is $(10)$.  Similarly, strong
transitions are $(10)\rightarrow(11)$ or $(10)\rightarrow (20)$.
$(10)\rightarrow(11)$ transition overlaps with the 0.7 keV feature.
$(10)\rightarrow(20)$ transition is irrelevant to the \pulsar spectrum
because it appears below the EPIC band ($E\lax0.3$ keV).  However, the
relative population of the $(10)$ state (1st excited state) to $(00)$
state (ground state) is $\sim\exp{(-\epsilon/kT)} \lax 10^{-2}$ where
$\epsilon \sim 1$ keV is the excitation energy and $kT\lax200$ eV. We
note that occupation probability of the excited states will be further
reduced by pressure ionization (\S\ref{sec_pbroad}). Therefore,
$(10)\rightarrow(11)$ transition will be much weaker than the
transitions from $(00)$ state in the 0.7 keV
feature. $(10)\rightarrow(02)$ transition is allowed in the
$\alpha=-1$ mode. However, transitions in the $\alpha=-1$ mode are in
general very weak hence we do not consider them further in this
section. We will present cross sections in the $\alpha=-1$ mode in
\S\ref{sec_cs} and they are included in our opacity calculations.

\subsubsubsection{He-like ion}

\begin{figure}
\epsscale{1.0} \plottwo{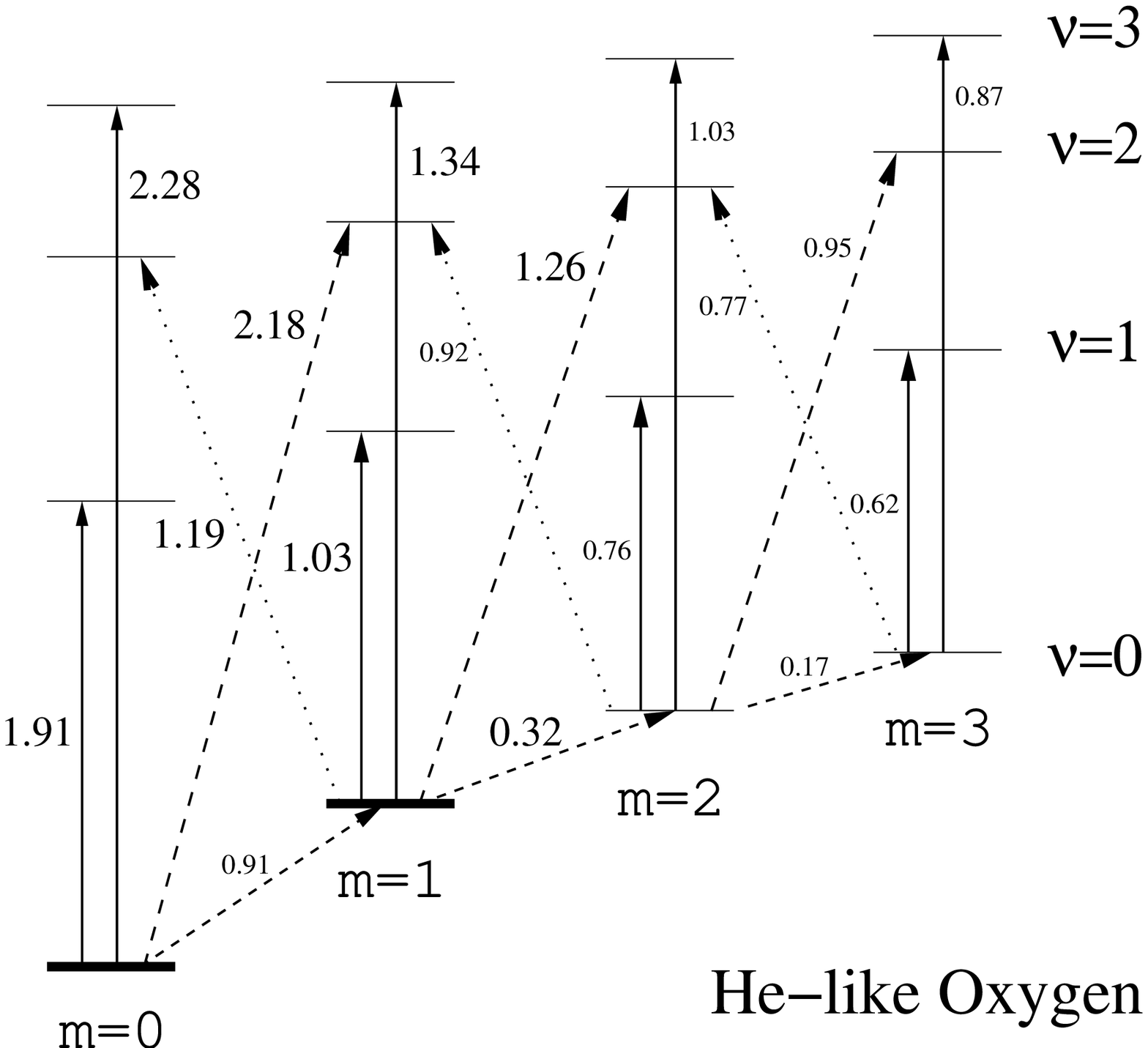}{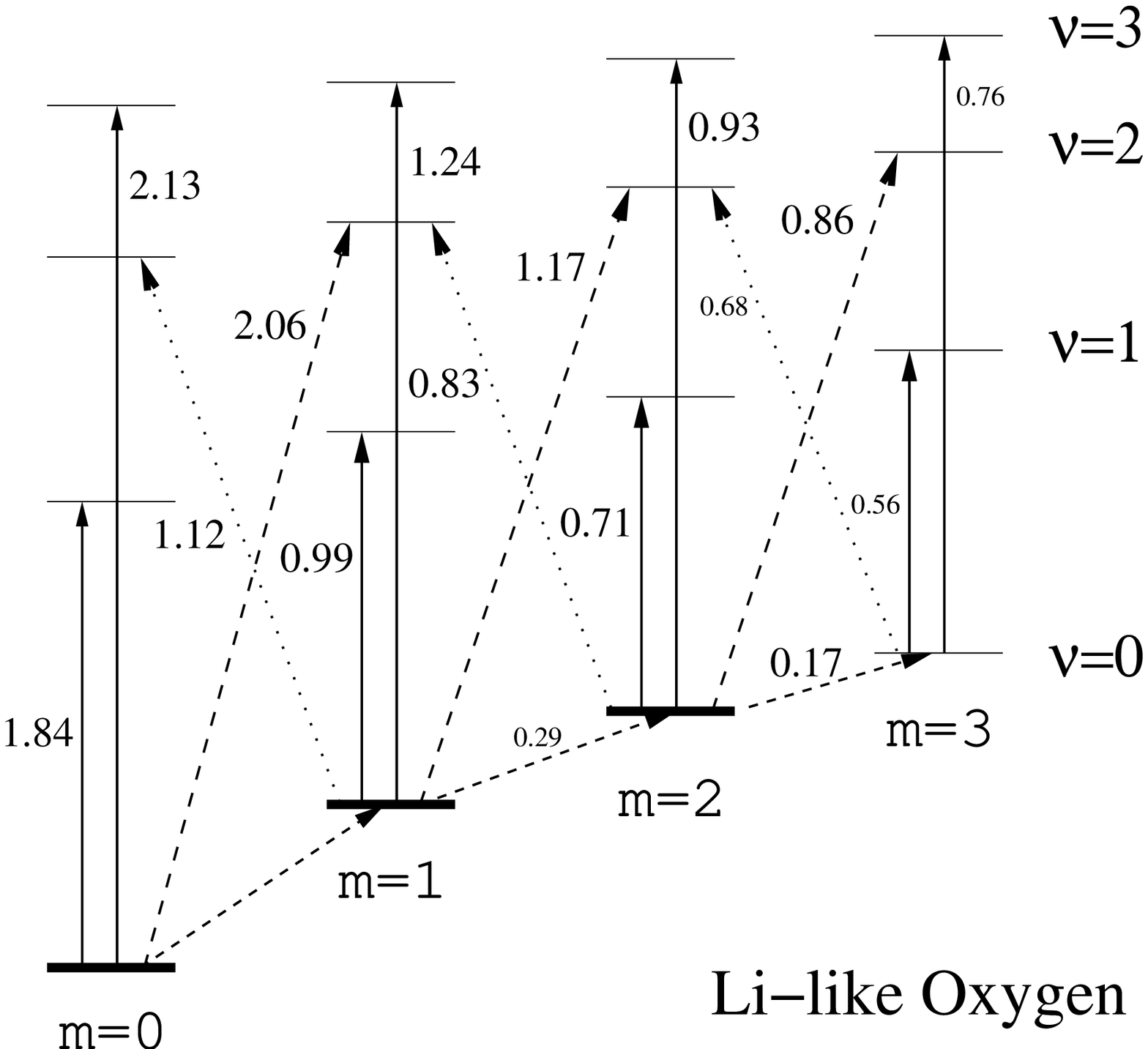}
\caption{Grotrian diagram for strong transitions of He-like (left) and Li-like
 (right) oxygen at $B=10^{12}$ G. See the caption of figure
 \ref{fig_lines} for notations. \label{fig_lines2}}
\end{figure}

\begin{figure}
\epsscale{1.0} \plottwo{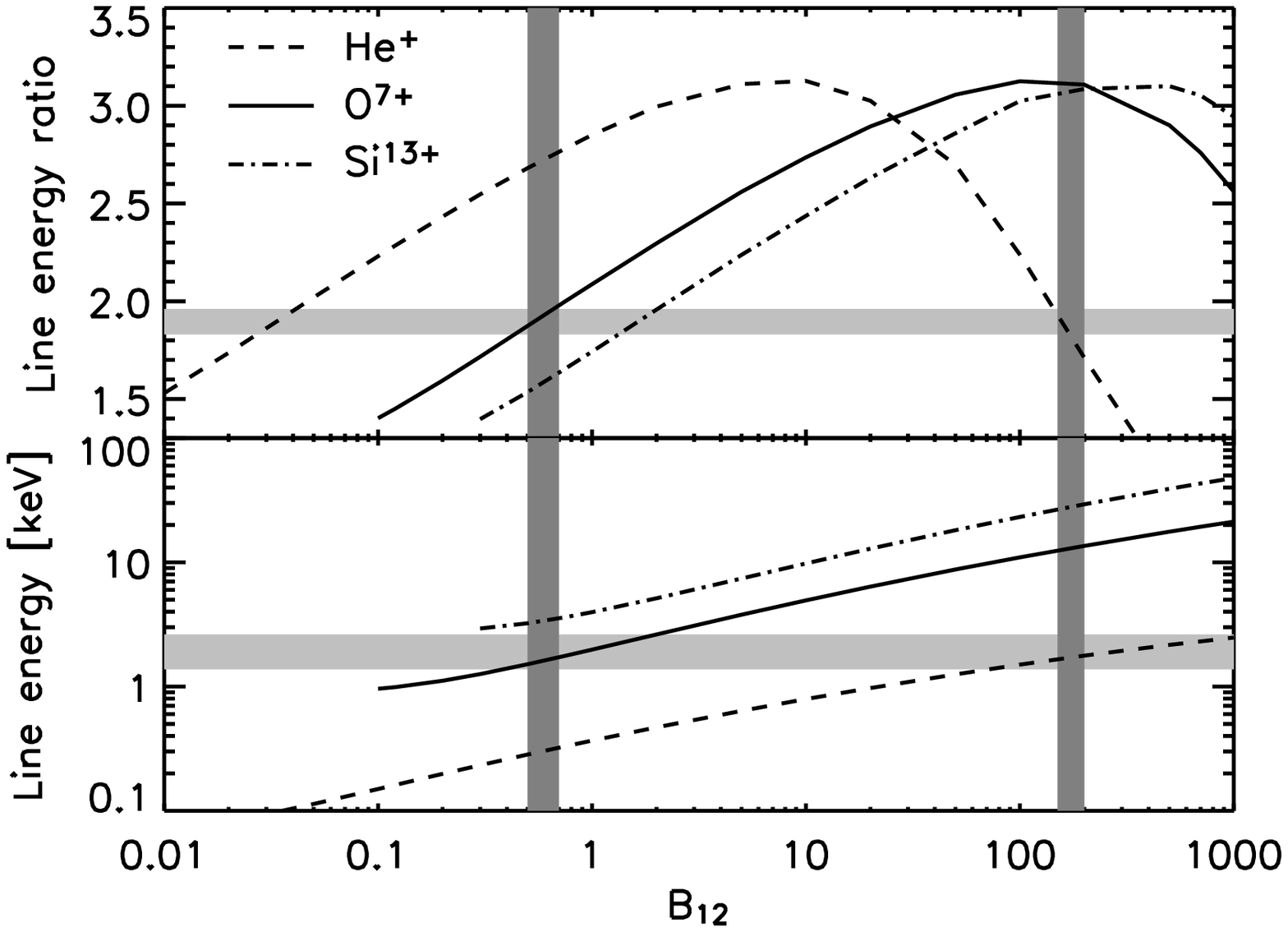}{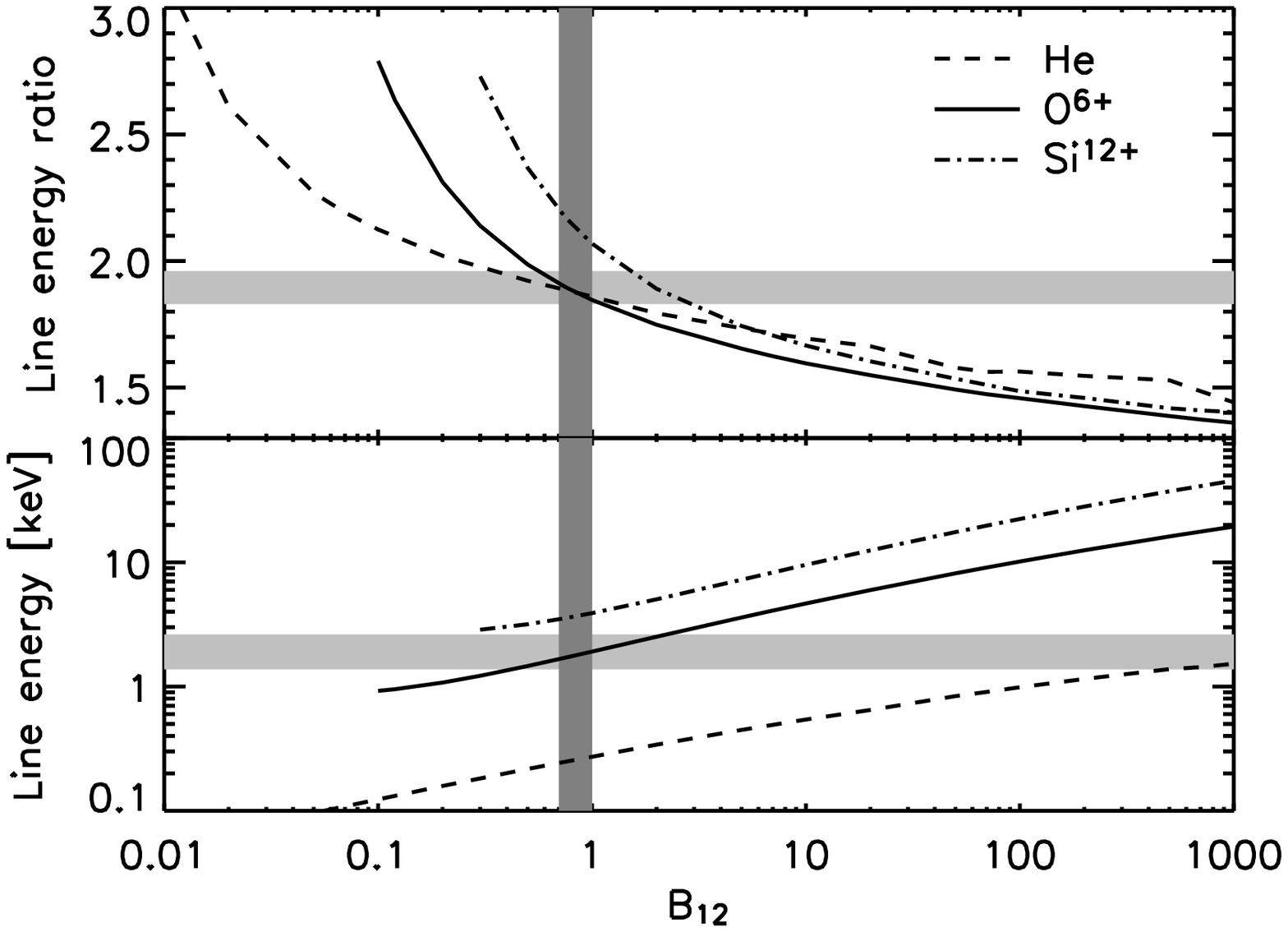}
\caption{Line energy ratio of H and He-like ions as a function of
magnetic field strength. The top figures present the line energy
ratios to illustrate condition (b). The bottom figures plot line
energies from $(00)\rightarrow(01)$ transition to illustrate condition
(c). We considered cases for helium (dashed line), oxygen (solid line)
and silicon (dotted-dashed line). For H-like ions, the transition is
$(00)\rightarrow(01)$ and $(00)\rightarrow(10)$. For He-like ions, the
transition is $(00)(10)\rightarrow(01)(10)$ and
$(00)(10)\rightarrow(00)(11)$. The grey areas indicate the measured
ratio and energy. Note that line energies are unredshifted and the
grey areas are obtained by the measured line energy multiplied by a
factor of $(1+z)$ with the range of $z=0$--$0.85$. The black area
shows the magnetic field range which satisfies both the line energy
ratio (top panel) and transition energy consistent with the \xmm data
(bottom panel). Cut-off at lower $B$ indicates that the MCPH$^3$ code
provides inaccurate results. \label{fig_ratio}}
\end{figure}

The ground state configuration of He-like ions is
$(00)(10)$.  See figure \ref{fig_lines2} for a Grotrian diagram of
He-like oxygen at $B=10^{12}$ G.  
Tight-loose transitions from the ground state such that
$(00)(10)\rightarrow(01)(10)$ and $(00)(10)\rightarrow(00)(11)$
overlap with the 1.4 and 0.7 keV feature
respectively. $(00)(10)\rightarrow(00)(20)$ transition appears below
the EPIC band. 

The 1st excited state of He-like ions is $(00)(20)$ state. Level
population relative to the ground state is $\lax0.2$ at
$kT\lax200$ eV. $(00)(20)\rightarrow(01)(20)$ transition overlaps with
the 1.4 keV feature. $(00)(20)\rightarrow(00)(21)$ transition appears
around 0.5 keV which may partially constitute the 0.7 keV
feature. $(00)(20)\rightarrow(10)(20)$ tight-tight transition overlaps
with the 0.7 keV feature. The other transitions from the 1st excited
state are either very weak or outside the EPIC band.

The 2nd excited state is $(00)(30)$ state with level population
$\lax0.1$ at $kT\lax200$ eV. Similarly,
$(00)(30)\rightarrow(01)(30)$ transition overlaps with the 1.4 keV
feature, while $(00)(30)\rightarrow(10)(30)$ transition overlaps with
the 0.7 keV feature. Transitions from other excited states are
unimportant in the X-ray band.

\subsubsubsection{Li-like ion}

The ground state configuration of Li-like ions is
$(00)(10)(20)$.  See figure \ref{fig_lines2} for a Grotrian diagram of
Li-like oxygen at $B=10^{12}$ G. $\Delta\nu=1$ transitions from $(00)$
and $(10)$ overlap with the 1.4 and 0.7 keV feature respectively.  Li-like ions
have an additional line due to $(20)\rightarrow(21)$ whose transition
energy is $\sim$0.5 keV which may merge with the 0.7 keV feature,
therefore it is consistent with the
data. $(00)(10)(20)\rightarrow(00)(10)(30)$ transition has too low
line energy to be observed in the X-ray band.  The 1st excited state
is $(00)(10)(30)$ with level population $\sim0.2$ relative to the
ground state. Similarly, tight-tight transitions to $(01)(10)(30)$ and
$(00)(11)(30)$ will appear in the 1.4 and 0.7 keV feature 
respectively. We have included other relevant transition lines in our
final results. 

\subsubsubsection{Be-like ion and other more neutral ions}

Ions with more than three electrons (e.g. Be-like ion) may not be
 plausible simply because additional unobserved features would appear
 in the EPIC energy band. For instance, Be-like ions will produce a
 fourth absorption line at $\sim 0.4$ keV due to the
 $(30)\rightarrow(31)$ transition. $(30)\rightarrow(40)$ transition is
 located below the EPIC band. Similarly, more neutral ions have
 $(00)\rightarrow(01)$ transition overlapping with the 1.4 keV
 feature, while other photo-absorption lines are located below 0.4
 keV.

\subsubsubsection{Relaxation of the single ionization state assumption
\label{sec_bbtransition}}

It is more plausible to interpret the two features as transitions from
ions with a few electrons (either H, He and Li-like ion) for the
following reasons. The separation in binding energies for two adjacent
states is largest for the $(00)$ and $(10)$ states and two adjacent
tightly-bound states at $m\ge2$ do not have a line energy ratio larger
than 2.  From the measured line energy ratio $\sim1.9$, we conclude
that these two features are from $(00)$ state for H-like ions, $(00)$
and $(10)$ states for He- and Li-like ions because they are the only
combination giving the right ratio. For Li-like ions, the 3rd
absorption line due to $(20)\rightarrow(21)$ transition may contribute
to the 0.7 keV feature. Be-like ions and more neutral ions will show
extra absorption features below 0.5 keV. In the next section, we
constrain $B$ and $Z$ based on the assumption that the two absorption
features are produced by transition lines from highly-ionized atoms.

\subsubsection{Constraining magnetic field strength from condition (b) \label{sec_b_const}} 

We constrain magnetic field strength ($B$) from the observation (b)
that the measured line energy ratios are $\sim1.9$.  As we discussed
in the previous section, the strong absorption lines relevant to the
two features are $(00)\rightarrow(01)$ and $(00)\rightarrow(10)$
transition for H-like ions and $(00)(10)\rightarrow(01)(10)$ and
$(00)(10)\rightarrow(00)(11)$ transition for He-like ions.  Figure
\ref{fig_ratio} shows line energy ratios for H-like and He-like ions
of helium, oxygen and silicon as illustrative examples. For H-like
ions, there are two solutions: oxygen at $\sim10^{12}$ G or helium at
$\sim10^{14}$ G. The latter corresponds to the model by
\citet{sanwal02} and \citet{pavlov05}
\footnote{At such high B-field, there is another quantum number
associated with ionic motion. \citet{pavlov05} attributed the 0.7 keV
feature to a transition with $\Delta m =1$ and an increment in the new
quantum number.}. On the other hand, there is only one magnetic field
value for a given line energy ratio for He-like ions. For Li-like
ions, the 3rd absorption line due to $(20)\rightarrow(21)$ transition
contributes to the lower energy line therefore B-field determination
will be more complicated.

\subsubsection{Constraining atmospheric composition from condition (c) \label{sec_z_const}} 

We constrain atmospheric composition ($Z$) from the observation (c)
that the absorption features appear at 0.7 and 1.4 keV. The lower
panel in figure \ref{fig_ratio} shows unredshifted transition energies
from the $(00)$ state for H-like ions and the $(00)(10)$ state of
He-like ions of helium, oxygen and silicon.  Elements above $Z=10$
(e.g. silicon) are excluded because transition energies are too high,
which requires unreasonably large gravitational redshift to place
lines in the EPIC band.

Searching over $(B,Z)$ phase space in the range of
$B=10^{11}$--$10^{14}$ G and $Z=1$--$26$ in the manner described
above, {\it the only candidates are oxygen and neon at $B\sim10^{12}$
G or helium at $B\sim10^{14}$ G}. Note that the Helium solution
requires that H-like helium be predominantly populated in the
photosphere, while the O/Ne solution allows a range of ionization
states. Later, we find the Helium solution implausible on the basis of
ionization balance and line strength (\S\ref{sec_helium}).  Table
\ref{tab_line} shows (unredshifted) line energies for H-, He- and
Li-like oxygen. In the Neon case, both magnetic field strength and
gravitational redshift are slightly higher than the Oxygen case. For a
given element and ionization state, gravitational redshift is
simultaneously determined. With the CCD energy resolution, we cannot
measure gravitational redshift with high accuracy nor distinguish
between Oxygen and Neon models. Here we give the range of the
gravitational redshift for each case; $B_{12}\simeq0.5-1,
z\simeq0-0.4$ for oxygen and $B_{12}\simeq1-2, z\simeq0.4-0.8$ for
neon.

%=============================================================================

\subsection{Case B: Pure cyclotron lines \label{sec_caseB}}

We consider the cases involving only cyclotron lines. Unlike case A,
here we simply present several possible solutions (see also table
\ref{tab_check}) and defer discussion on their plausibility to
\S\ref{sec_others}.

\subsubsection{Electron cyclotron line \label{sec_ecyclo}} 

The fundamental and 2nd harmonic of electron cyclotron lines have a
line energy ratio of two \citep{xu03, bignami03, deluca04}. Magnetic
field strength is $\sim 8\times10^{10}$ G to match the observed line
energies in \pulsar.

\subsubsection{Ion cyclotron line \label{sec_icyclo}} 

When $B\uax10^{14}$ G, ion cyclotron line energy is in the X-ray band.
Unlike electron cyclotron lines, ion cyclotron line energy depends on
ionization state and atomic mass ($\hbar\omega_i = 6.30(Z_i/A)B_{12}$
[eV] where $Z_i=Z-n_e$ and $A$ is atomic number). There are three
possible solutions associated with ion cyclotron lines; (1) the
fundamental and 2nd harmonics, (2) the fundamental lines from two
different elements with atomic mass differ by 2 (e.g. H and He), (3)
the fundamental lines from two different charge states (e.g. bare and
H-like Helium ion).

%===========================================================================

\subsection{Case C: mixture of atomic transition and cyclotron line \label{sec_caseC}}  

We consider cases in which one of the features is a cyclotron line and
the other is an atomic transition line. The ionization state must be
hydrogenic since otherwise unobserved features from less ionized
states will deviate from the \xmm spectra. As discussed in case A, the
feasible atomic transition is either $(00)\rightarrow (10)$ or
$(00)\rightarrow (01)$ transition line of H-like ions.

\subsubsection{Electron cyclotron line and atomic transition \label{sec_mix1}} 

When the 1.4 keV feature is an electron cyclotron line
($B\sim2\times10^{11}$ G), the 0.7 keV feature must be due to either
$(00)\rightarrow (01)$ or $(00)\rightarrow(10)$ transition of H-like
ion.  In either case, the solution requires H-like nitrogen or oxygen.

When the 0.7 keV feature is an electron cyclotron line
($B\sim8\times10^{10}$ G), our atomic data is inaccurate for $(00)$
states of mid-Z elements. Hence, we estimated line energies from the
atomic data of \citet{ruder} for Hydrogen atom and a well-known
scaling law with $Z$ \citep{ruder}. We found that it requires H-like
ions of $Z=10$--12.

\subsubsection{Ion cyclotron line and atomic transition \label{sec_mix2}} 

In this case, it requires $B>10^{14}$ G to match an ion cyclotron line
with one of the two features. The only solution is that the 0.7 keV
feature is an ion cyclotron line and the 1.4 keV feature is either
$(00)\rightarrow(01)$ transition of H-like helium or
$(00)\rightarrow(10)$ transition of H-like lithium at
$B\sim2\times10^{14}$ G. The former case is nearly identical to the
Helium model proposed by \citet{sanwal02}.

%=============================================================================

\section{Oxygen atmosphere model at B$=10^{12}$ G and $z=0.3$ \label{sec_oxygen}} 

In this section, we calculate LTE opacities for Oxygen atmosphere at
$B=10^{12}$ G and $z=0.3$. $z=0.3$ was chosen so that the calculated
line energies roughly match the observed feature energies. We note
that all the figures in this section are presented in the observer's
frame after correcting line energies by the redshift factor. We note
that the Neon case produces almost identical line energies and
opacities with slightly larger $B$ and $z$ values.

\subsection{Cross sections \label{sec_cs}} 

\begin{figure}
\epsscale{0.7} \plotone{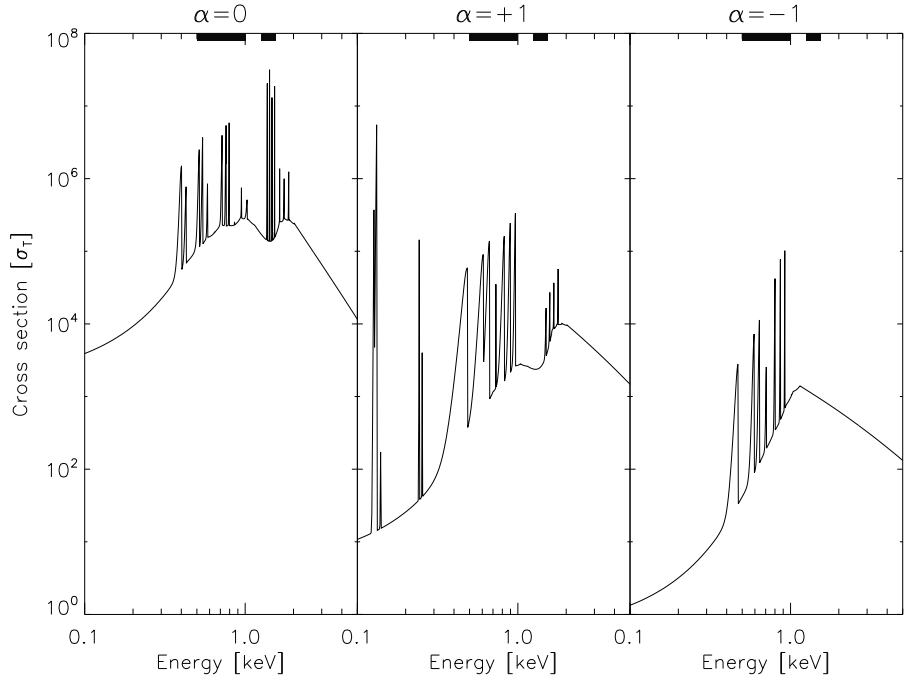}
\caption{Bound-bound and bound-free transition cross sections of
H-, He-, Li- and Be-like oxygen for $\alpha=0$ (left), $\alpha=+1$ (middle) and $\alpha=-1$ (right) polarization mode at $B=10^{12}$ G. Photon energies are shifted by $(1+z)=1.3$. The observed feature energies are indicated by black boxes. All the ions are assumed to have the same ionization fraction.  
For transition lines from excited states, we multiplied oscillator 
strengths by the Boltzmann factors assuming $kT=150$ eV. \label{fig_cs}}
\end{figure}

We calculated cross sections for bound-bound transitions and
bound-free transitions of all the ionization states of oxygen. Figure
\ref{fig_cs} shows the cross sections of H-, He-, Li- and Be-like
oxygen since we will see later that they are the dominant ionization
states in the \pulsar atmosphere. They are presented for the three
cyclic polarization modes ($\alpha=0, \pm1$) and photon energies are
shifted by $(1+z)=1.3$ to roughly match the observed feature energies
(indicated by black boxes in the figure). The lines are broad and
asymmetric with low energy wings due to motional Stark effects (see
\S\ref{sec_ms}). In all the polarization modes, bound-free transitions
have much smaller cross sections than bound-bound transitions. The
bound-free cross sections are relatively small partially because
continuum states are subject to significantly larger motional Stark
broadening \citep{pavlov93}. The 1.4 keV feature is mainly composed of
$\alpha=0$ transitions along with several weak $\alpha=+1$
transitions. In all the polarization modes, there are a large number
of transition lines in 0.5--1.0 keV. Less-ionized oxygen will have
more bound-bound transitions below 0.5 keV. We also included the
free-free absorption in our model from \citet{potekhin03}. However,
the free-free absorption is unimportant in the X-ray band because the
cross sections are significantly smaller than those of bound-bound and
bound-free absorption and oxygen is mostly partially-ionized.

\subsection{Ionization balance \label{sec_density}} 

In this section, we investigate ionization balance of an Oxygen
atmosphere at $B=10^{12}$ G. In LTE atmospheres, degree of ionization
and level population are determined by the Saha-Boltzmann
equilibrium. The generalized Saha equation in the presence of a
magnetic field is given by \citep{khersonskii87, rajagopal97},
\begin{equation}
\frac{n_i}{n_{i+1}n_e}=\frac{1}{2}\left(\frac{2\pi\hbar^2}{m_e
kT}\right)^{3/2}\frac{\tanh{\eta_e}}{\eta_e}\frac{\eta_{i}}{\sinh{\eta_{i}}}\frac{\sinh{\eta_{i+1}}}{\eta_{i+1}}e^{\chi_i/kT}\frac{Z_{i}}{Z_{i+1}},
\label{eq_saha}
\end{equation} 
where $n_i$ is number density of an ionization state $i$,
$\eta_e=\hbar\omega_e/2kT$ and $\eta_i=\hbar\omega_i/2kT$. $\chi_i (>0)$ is
the ionization energy. $Z_{i}$
is the internal partition function (IPF) reflecting level population
of bound electrons of an ionization state $i$.

In a high density plasma, electron bound states are destroyed (or
depopulated) by the electric field from adjacent ions (Pressure
ionization). Pressure ionization can be taken into account by
assigning occupation probabilities (OP) for bound states
\begin{equation}
Z_{i}=\sum_\kappa w_{i,\kappa} g_{i,\kappa} \exp{(-\epsilon_{i,\kappa}/kT)},
\end{equation} 
where $w_{i,\kappa}$ is the occupation probability and $g_{i,\kappa}$ is the degree of
degeneracy of a bound state $\kappa$ of an ionization state
$i$. $\epsilon_{i,\kappa} (>0)$ is the excitation
energy. Among various approaches for
evaluating the OP, we adopted the formalism of \citet{potekhin02}. We
used the fitting formula derived by \citet{potekhin02} which reproduce
the electric microfield distributions from their Monte-Carlo methods to 
within a few percent. The microfield distribution of 
\citet{potekhin02} is highly accurate over a large range of the plasma
coupling constant ($\Gamma = 0$--100). It fully covers the $(\rho, T)$
space discussed in \S\ref{sec_param_space}. We 
note that the method of \citet{potekhin02} also takes into account
electron screening of the Coulomb field.

\begin{figure}
\epsscale{1.0} \plottwo{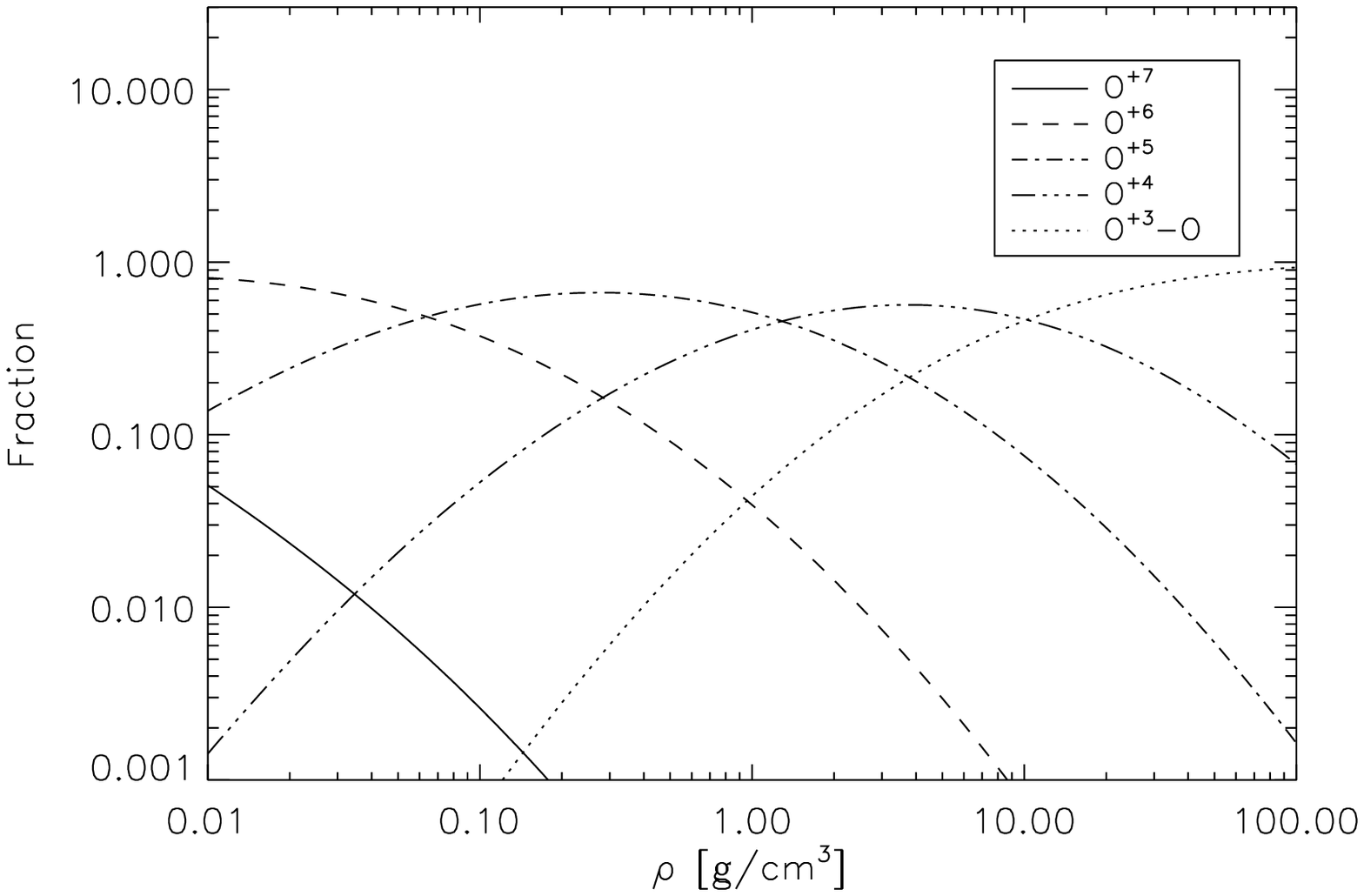}{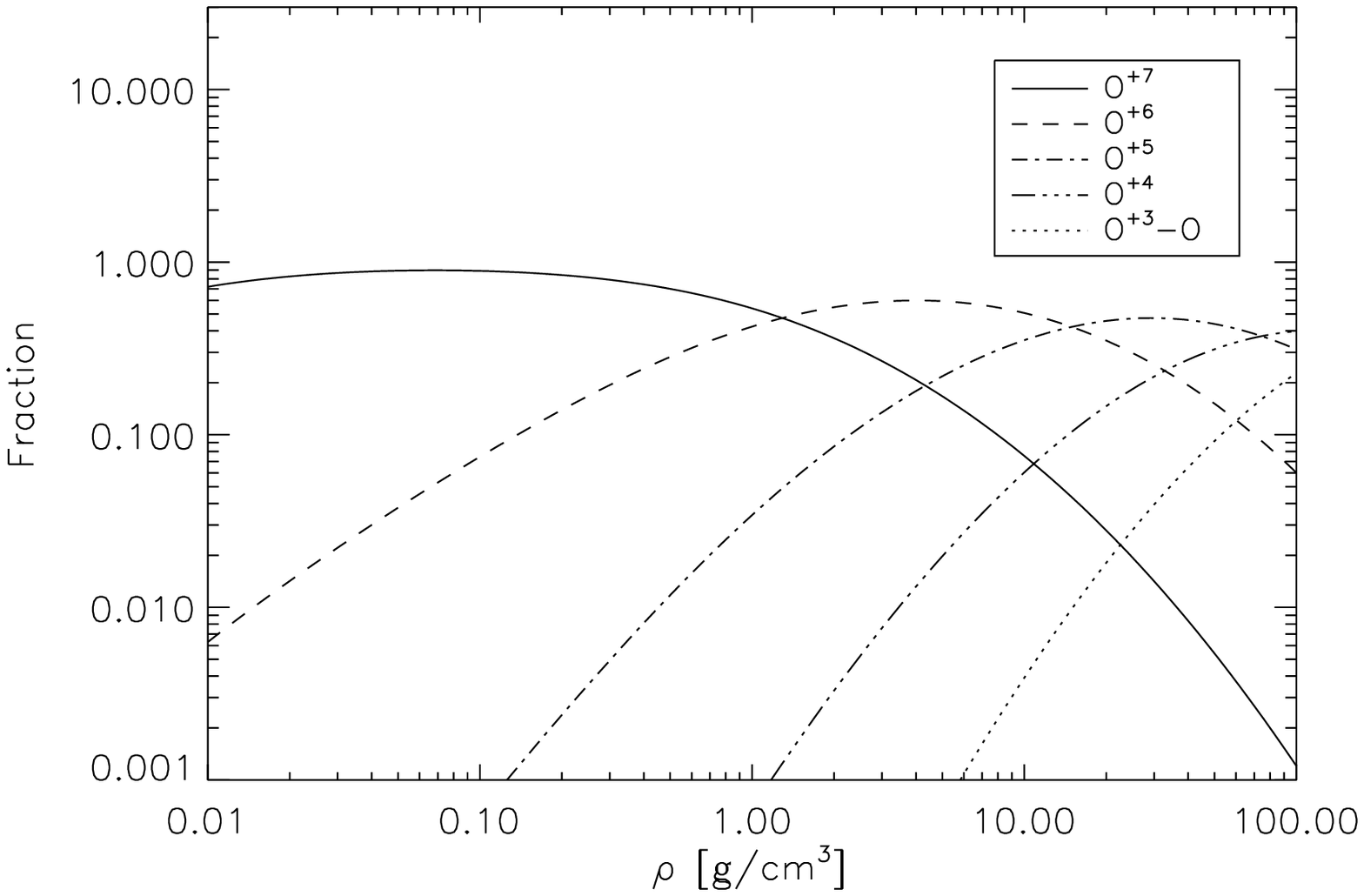}
\caption{Ionization balance of oxygen at $B=10^{12}$ G and $kT=100$ eV
(left) and $kT=200$ eV (right). The dotted curve shows the summed ionization fraction from B-like to neutral oxygen. \label{fig_ionization}}
\end{figure}

\begin{figure}
\epsscale{0.7} \plotone{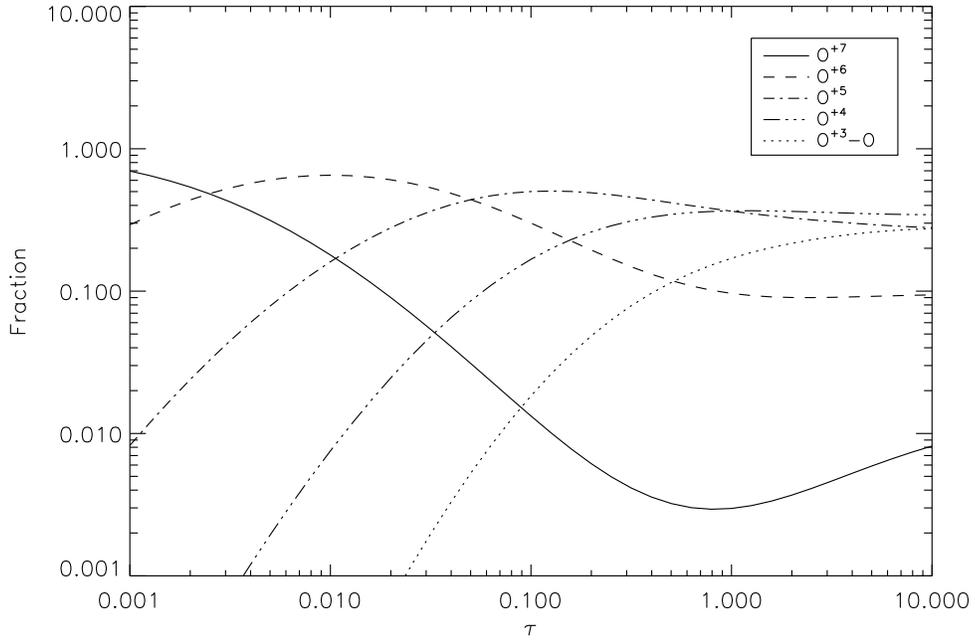}
\caption{Ionization fraction of oxygen for the grey temperature profile at $kT_{eff}=200$ eV and $B=10^{12}$ G. The dotted curve shows the summed ionization fraction from B-like to neutral oxygen. \label{fig_grey}}
\end{figure}

We computed the IPF for each ionization state of oxygen using binding
energies calculated by the MCPH$^3$ code. We iteratively solved the
coupled Saha equations for different charge states until convergence
was achieved.  Figure \ref{fig_ionization} shows the fraction of
various ionization states of oxygen for $kT=100$ eV and 200 eV at
$B=10^{12}$ G.  H, He and Li-like ions for both the Oxygen and Neon
case are largely populated at $\rho\sim10^{-2}$--$10^0$ g/cm$^{3}$ at
$kT=100$ eV and $\rho\sim10^{-2}$--$10^{2}$ g/cm$^{3}$ at $kT=200$
eV. These temperature and density ranges are typical of the
photosphere of \pulsar. To be more realistic, we have calculated
ionization fraction for a grey temperature profile at $kT_{eff}=200$
eV. \citet{miller92} and \citet{rajagopal97} showed that mid-Z
element atmospheres have nearly grey temperature profile contrary to
Hydrogen or Helium atmospheres because bound-bound and bound-free
opacities dominate in mid-Z element atmospheres. Figure \ref{fig_grey} shows
ionization fraction as a function of optical depth (in units of
Thompson depth). Highly-ionized oxygen is largely populated at
$\tau=10^{-4}$--$10^{0}$ where the absorption lines are likely
formed. 

%--------------------------------------------------------------------------

\subsection{Line broadening \label{sec_broad}}

We investigated various line broadening mechanisms in NS
atmospheres. They are compared with the observation (d) that the 0.7
keV feature ($\Delta E/E \sim 40$\%) has larger line width than the
1.4 keV feature ($\Delta E/E \sim 15$\%). In the opacity calculations,
we included all the line broadening effects except magnetic field
variation over the line-emission area (hereafter we call it magnetic
broadening for convenience) as its distribution is
unknown. Nevertheless, we estimated the degree of magnetic broadening
in comparison with the \xmm observation. The results are summarized in
table \ref{tab_broad}.

\subsubsection{Doppler broadening \label{sec_doppler}} 

For the temperature and spin period of \pulsar, thermal and rotational
Doppler broadening are negligible ($\Delta E/E < $ 0.1\%).

\subsubsection{Pressure broadening \label{sec_pbroad}}

In a dense plasma, electron bound states are perturbed by electric
fields from electrons and ions \citep{salzmann98}. The former is
caused by the interaction of a radiator with fast free electrons
(electron collisional broadening) while the latter is due to the
electric field of adjacent ions (quasi-static Stark
broadening). Higher excited states are subject to larger
broadening. Following \citet{rajagopal97}, we estimated that pressure
broadening produces line broadening of less than 1\% for transitions
involving $\nu=1, 2$ states even at the highest photosphere density
($\rho\sim10^2$ g/cm$^3$). Pressure broadening is not responsible for
the broad absorption features.

\subsubsection{Motional Stark effects \label{sec_ms}}

As ions move randomly in a thermal plasma, coupling of the collective
motion and the internal electronic structure induces a motional Stark
field ($\vec{E}=\frac{\vec{v}\times\vec{B}}{c}$) in the ion's comoving
frame where $\vec{v}$ is the velocity of ion (i.e. thermal
velocity). Since the motional Stark field reduces the binding energy,
the line profile has a redward wing \citep{pavlov93}. Line width is
 estimated as
\begin{equation}
\Delta E_{MS} = kT(1-M_{\perp1}/M_{\perp2}),
\end{equation}
where $M_{\perp1}$ and $M_{\perp2}$ are the so-called anisotropic mass
of a ground state and an excited state defined in
\citet{pavlov93}. For each bound state, we evaluated $M_{\perp}$ using
the perturbation method adopted by \citet{pavlov93}.  The use of the
perturbation method is justified in our case because the atomic mass
is much heavier and the binding energies are significantly larger
compared to the energy shift caused by the motional Stark field. The
estimated line widths are $(10^{-2}-10^{-1})kT$ ($\Delta E/E \lax$ few \%). 

\subsubsection{Different ionization states \label{sec_blend}} 

Lines from different ionization states are likely to contribute to the
two broad features because of a temperature gradient in the
photosphere.  We estimated that line broadening due to different
transition lines can be as large as 20\% for both features. The 0.7
keV feature will have larger line width than the 1.4 keV feature
because $(00)\rightarrow(01)$ transition energies do not vary with
ionization states as the transition involves the innermost
electron. On the other hand, the 0.7 keV feature consists of many
transition lines from different ionization states. This is consistent
with the observation (d).

\subsubsection{Non-uniform surface magnetic field \label{sec_nonuniform}}

The non-uniformity of a surface magnetic field causes line broadening
and line energy shift with spin phase since the binding energy is a
function of magnetic field. The binding energy of tightly-bound states
varies with B-field logarithmically, while the binding energy of
loosely-bound states varies even more weakly \citep{ruder}. Therefore, the
degree of line broadening is larger for tight-loose transitions. For
instance a B-field variation by a factor of 2 (corresponding to the
magnetic field difference between the magnetic pole and\ equator in a
dipole configuration) will produce $\sim$ 30\% and $\sim 20$\% line
broadening for tight-loose and tight-tight transitions. However,
magnetic broadening alone cannot be responsible for the broad features
because the predicted line widths are comparable in both
features. Note that phase-resolved spectroscopy can decouple magnetic
broadening from other mechanisms since we see different parts of the
line-emitting region.

%------------------------------------------------------------------------

\subsection{Polarization vectors \label{sec_pol}}

In a strongly-magnetized plasma, radiation propagates in two normal
modes: an ordinary mode (O-mode) parallel to the B-field and an
extraordinary mode (X-mode) perpendicular to the B-field
\citep{ginzburg70, meszaros92}. The modal description of photon
propagation is valid in typical conditions of NS atmosphere except at
the resonance energies (e.g. vacuum resonance). Because the X-mode
opacities are smaller than the O-mode opacities, X-mode photons
decouple from matter at a deeper layer in the atmosphere where the
temperature is higher. Therefore, X-ray flux is predominantly carried
by the X-mode photons \citep{shibanov92}.

\begin{figure}
\epsscale{1.0} \plottwo{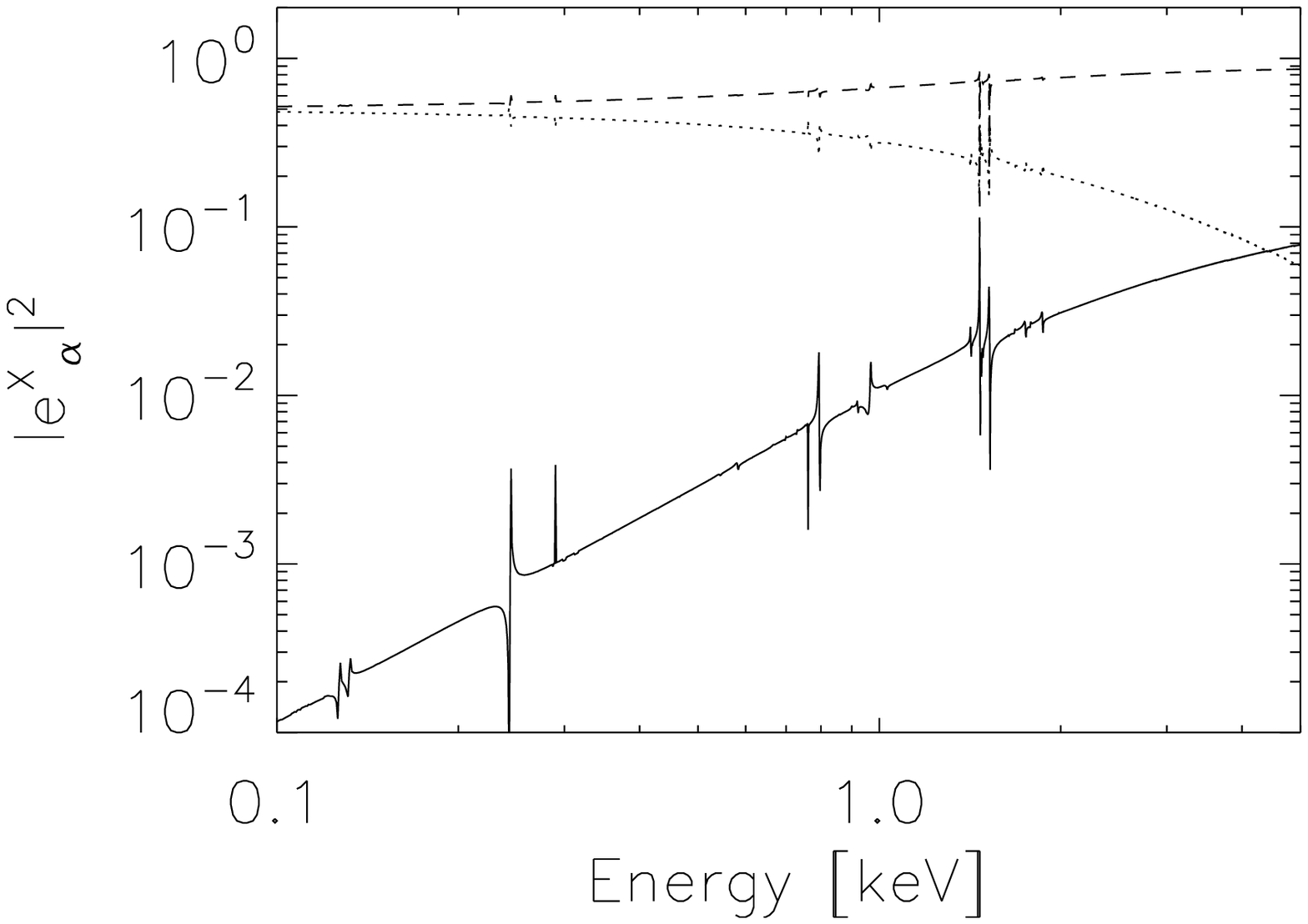}{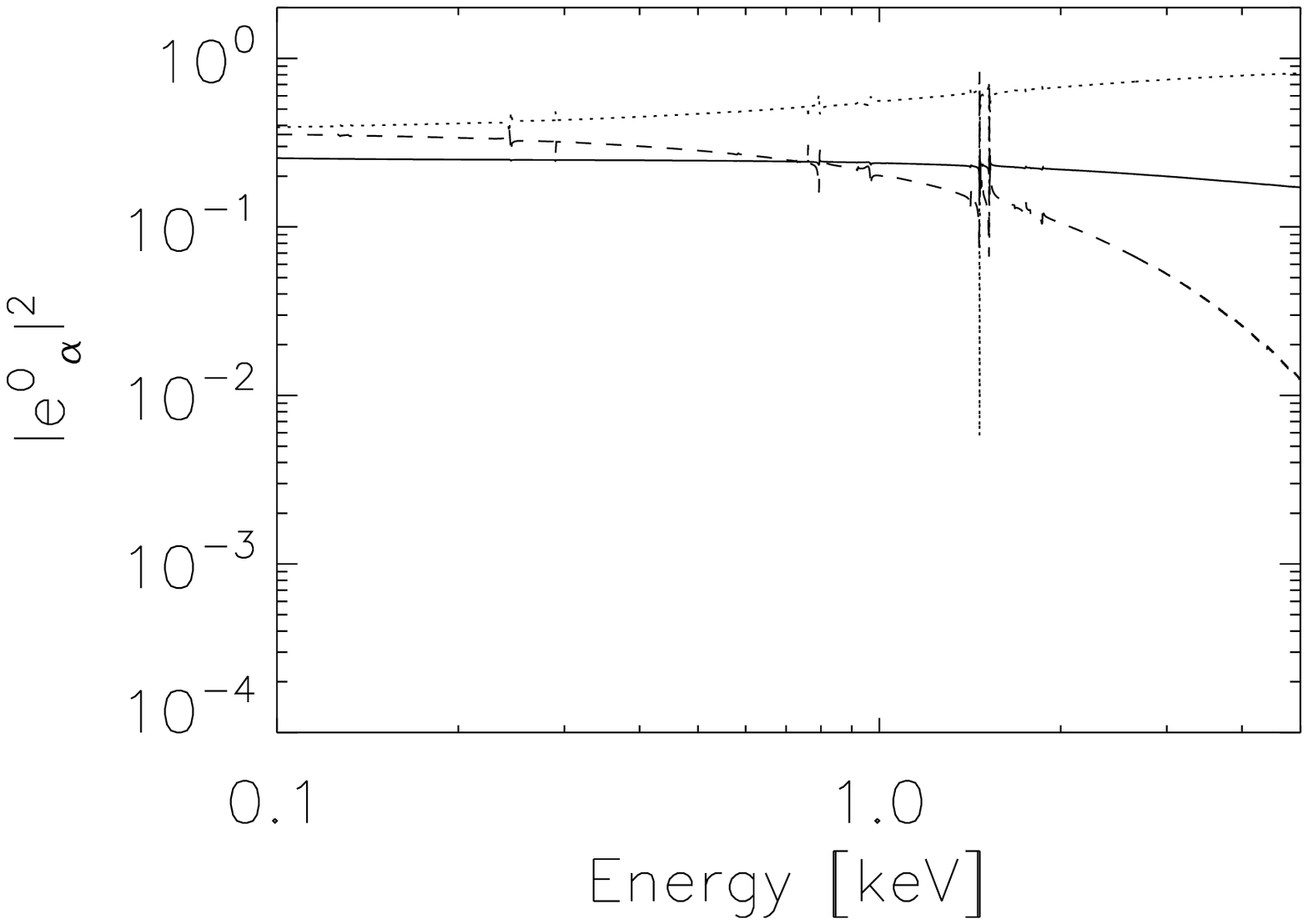}
\caption{Squared amplitude of the basic polarization vectors for
X-mode (left) and O-mode (right) at $\theta_B=30^\circ$, $\rho=0.1$
g/cm$^3$ and $kT=150$ eV. Solid, dashed and dotted lines indicate longitudinal
($\alpha=0$), right-circular ($\alpha=+1$) and left-circular ($\alpha=-1$) polarization respectively. The spikes are due to the resonant behavior of the
dielectric tensors at the transition
energies. \label{fig_polarization}}
\end{figure}

We calculated the polarization vectors ($\vec{e}_\alpha^j$ in equation
(1)) following \citet{bulik96} and \citet{potekhin04_2}. As the Oxygen
atmosphere is not fully-ionized, we included effects of the bound
species by use of the Kramers-Kronig relation and the absorption
coefficients we calculated. Figure \ref{fig_polarization} shows
$|e^{j}_\alpha (E)|^2$ for X-mode and O-mode at
$\theta_B=30^\circ$, $\rho=0.1$ g/cm$^3$ and $kT=150$ eV. In X-mode, the $\alpha=\pm1$ vector component is
about 1--2 orders of magnitude larger than the $\alpha=0$ vector
component over a large range of photon energy and $\theta_B$ (photon
propagation angle relative to magnetic field). At the resonance
energies, spikes are seen due to the large absorption coefficients. We
note that even near the resonance energies the modal description is
still valid since the polarization vector amplitudes of different basic
polarization modes do not cross each other.

\subsubsection{Vacuum resonance effects} 

We briefly describe the effects of vacuum polarization on spectral
features as it is important for several proposed models and the few
models involving ion cyclotron lines discussed in \S\ref{sec_method}
since they require $B>10^{14}$ G. When $B\uax B_C$ where
$B_C=4.414\times10^{13}$ G, vacuum polarization due to virtual
e$^+$e$^-$ pairs becomes important. In the photosphere, the effects of
the plasma and vacuum polarization become comparable (vacuum
resonance) and the two normal modes can switch with each other. The
net result is that the X-mode photosphere (where the X-mode photons
decouple from matter) is shifted upward in the atmosphere where
temperature is lower. As there is less temperature gradient between
the X-mode photosphere and the line-forming depth, spectral features
are strongly suppressed. Recent Hydrogen atmosphere models with full
radiative transfer solutions showed that both ion cyclotron lines and
atomic features become unobservable due to the vacuum resonance
effects \citep{ho03}. We note that the vacuum polarization is
negligible for an Oxygen atmosphere at $B\sim10^{12}$ G.

%========================================================================

\subsection{LTE opacities \label{sec_summary}}

In this section, we present the LTE opacities including all
the NS atmosphere physics addressed so far. We fixed $B=10^{12}$ G and
$kT=150$ eV to illustrate the density and angular dependence of the
opacities. Hereafter we focus on the X-mode opacities since 
the X-ray flux is predominantly carried by the X-mode photons at
$B\sim10^{12}$ G. 

\begin{figure}
\epsscale{1.0} \plottwo{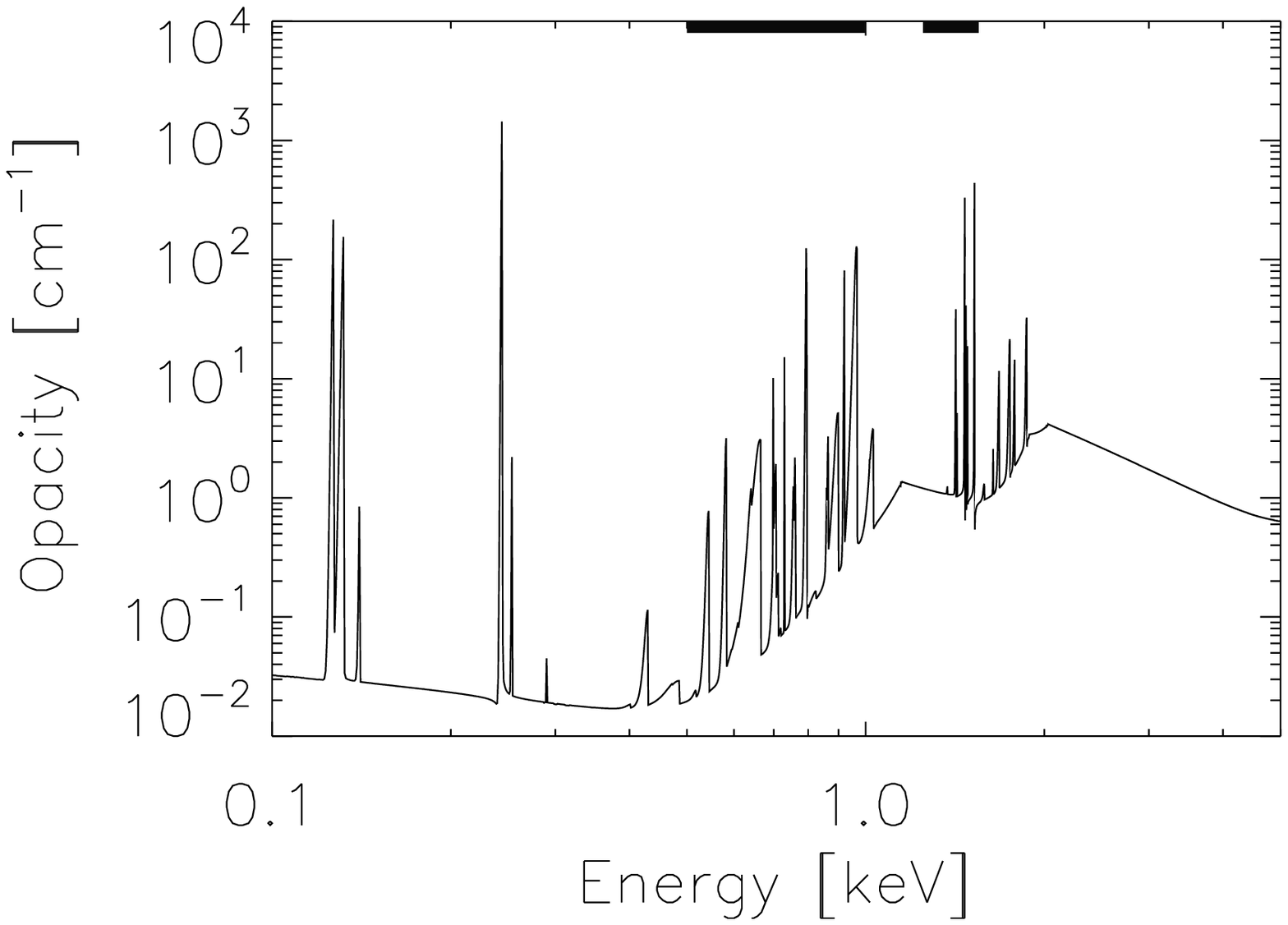}{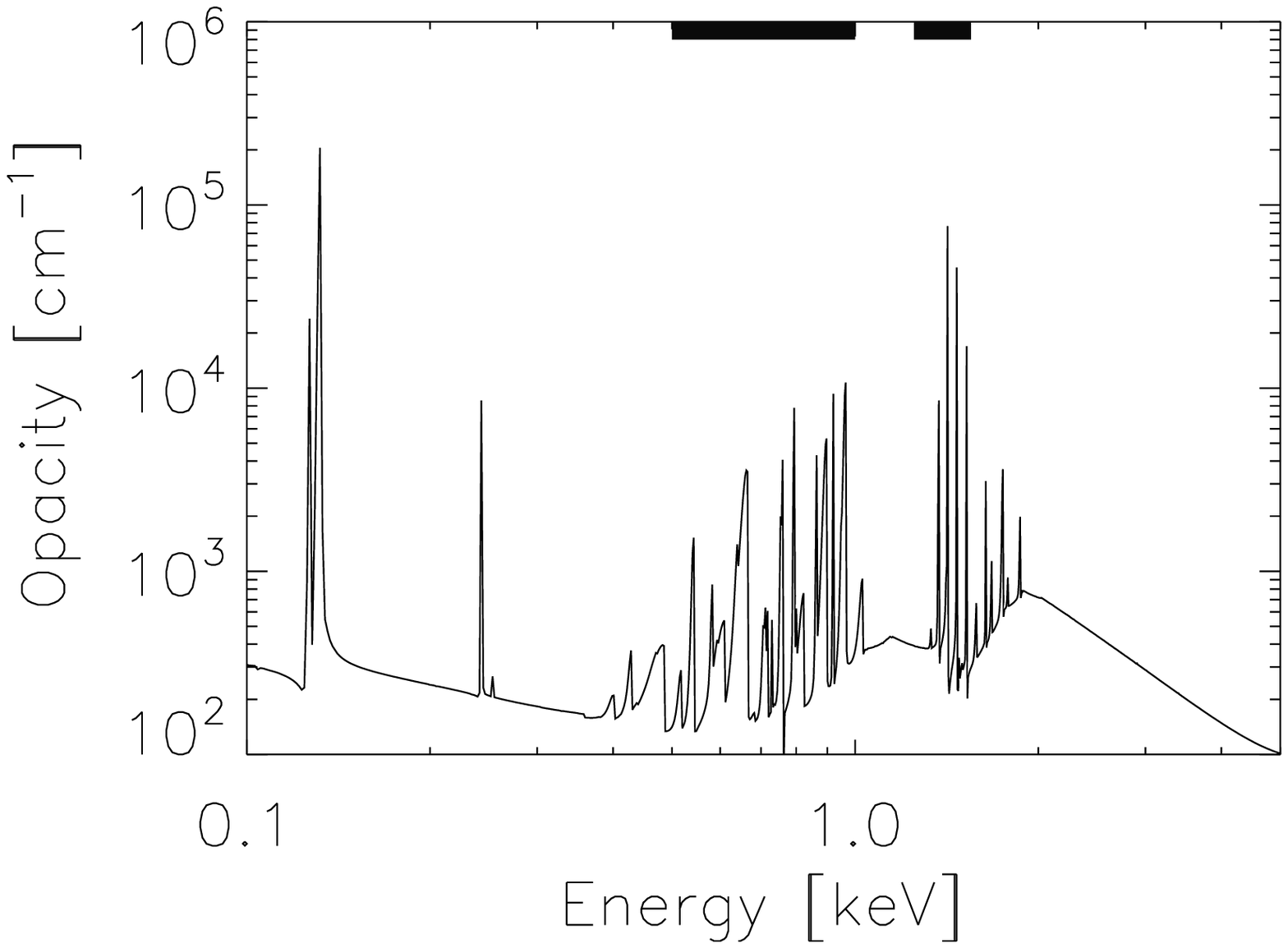}
\caption{X-mode opacities of oxygen at $\rho=0.1$ g/cm$^3$ (left)
and $\rho=10$ g/cm$^3$ (right). The other parameters are $B=10^{12}$
G, $kT=150$ eV and $\theta_B=30^\circ$. \label{fig_opacity_x}}
\end{figure}

Figure \ref{fig_opacity_x} shows X-mode opacities at $\rho=0.1$ 
g/cm$^3$ (left) and $\rho=10$ g/cm$^3$ (right) respectively 
($\theta_B=30^\circ$). 
At $\rho=0.1$ g/cm$^3$, $(00)\rightarrow(10)$ transition of H-like
oxygen is most prominent in the 0.7 keV feature while the 1.4 keV
feature is composed of $(00)\rightarrow(01)$ transition lines of
H-like and He-like oxygen. At $\rho=10$ g/cm$^3$,
$(00)\rightarrow(01)$ transitions of He- and Li-like ions are
strongest in the 1.4 keV feature. Unlike the $\rho=0.1$ g/cm$^3$
case, the 0.7 keV feature consists of tight-loose transition lines of
He- and Li-like ions.

\begin{figure}
\epsscale{1.0} \plotone{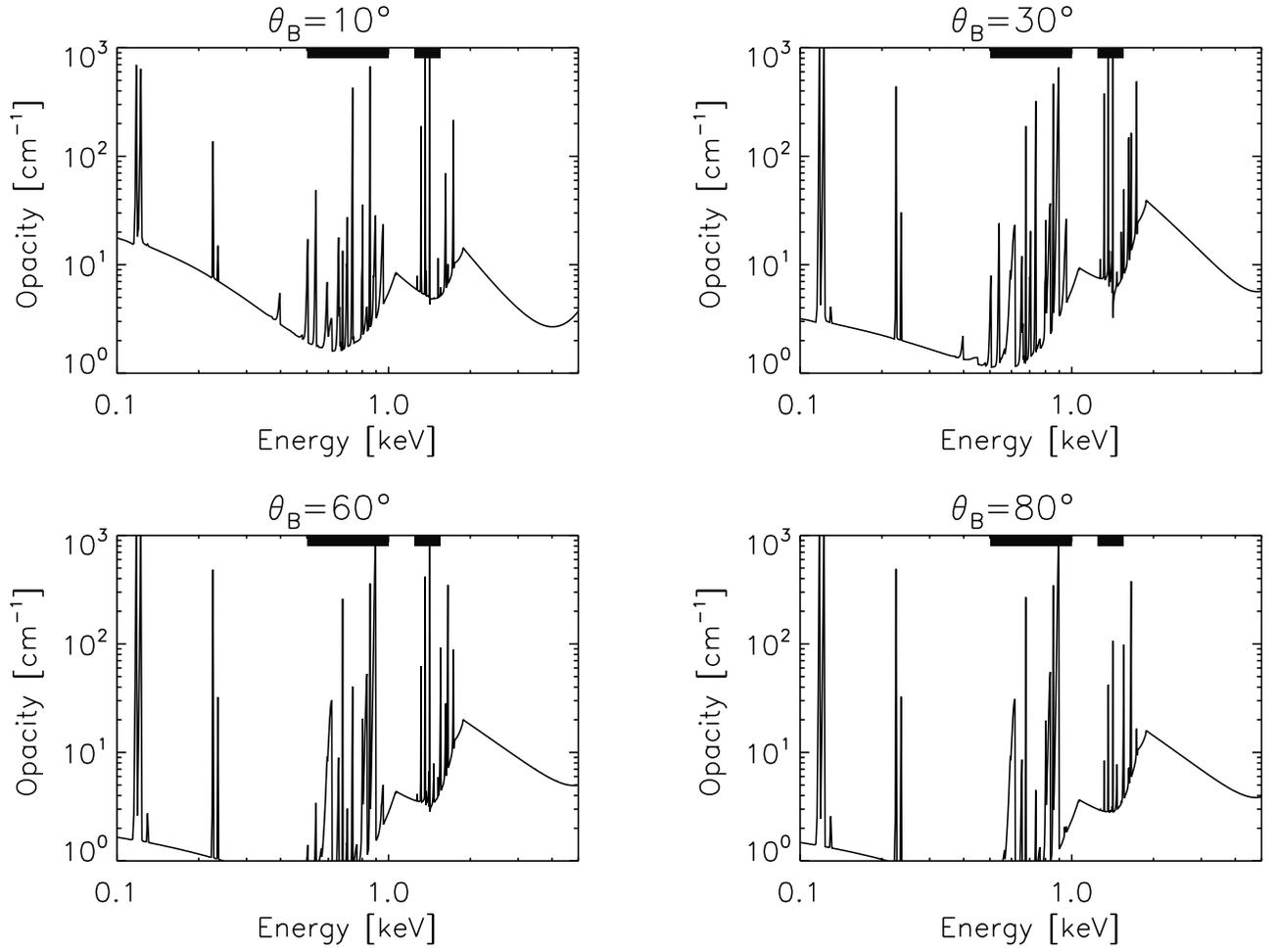}
\caption{X-mode opacities of oxygen at four different angles
($\theta_B=10, 30, 60$ and 80$^\circ$). The other parameters are
$B=10^{12}$ G, $kT=150$ eV and $\rho=1$
g/cm$^3$. \label{fig_angle}}
\end{figure}

\begin{figure}
\epsscale{0.6} \plotone{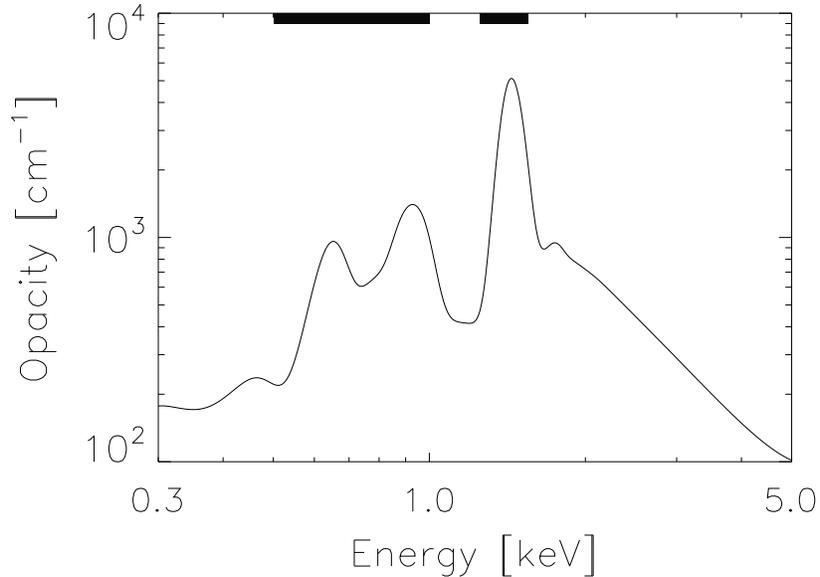}
\caption{X-mode opacities in figure \ref{fig_opacity_x} ($\rho=10$ g/cm$^3$) blurred by
\xmm/EPIC energy resolution function. \label{fig_ccd}}
\end{figure}

Figure \ref{fig_angle} shows the X-mode opacities at four different
angles. The other parameters are set to the same values ($\rho=1$
g/cm$^3$, $kT=150$ eV and $B=10^{12}$ G) to illustrate the angular
dependence. Note that the 1.4 keV feature varies with $\theta_B$ more
than the 0.7 keV feature because the 1.4 keV feature consists mainly
of tight-loose ($\alpha=0$) transitions and their polarization vectors
have strong angular dependence.  Nevertheless, over a large range of
$\theta_B$, three complex structures are seen at $\sim$ 0.2, 0.7 and
1.4 keV. These features are robust as they appear over a large range
of $\rho, T$ and $\theta_B$.  The feature at $\sim0.2$ keV is
unobservable by \xmm because the continuum flux is very low at
$E\sim0.2$ keV due to the high ISM neutral Hydrogen absorption. On the
other hand, the other two complex features nicely reproduce the two
discrete spectral features at $\sim0.7$ and $\sim1.4$ keV. They
consist of several narrow features, but they may be blurred by the
poor CCD resolution.  In figure \ref{fig_ccd} we show the X-mode
opacities convolved with the EPIC-PN energy resolution function
($\Delta E/E \sim 5-10$ \% \citep{xmm}). The 0.7 keV feature has
larger line width than the 1.4 keV feature. This is consistent with
the observation (d). Magnetic broadening will further broaden both
features by roughly the same degree (\S\ref{sec_nonuniform}).
Interestingly, the 0.7 keV feature shows substructure, while the 1.4
keV feature appears as a single broad Gaussian-like line. Our spectral
analysis of the 260 ksec \xmm data shows a possible substructure in
the residuals around 0.7 keV (figure \ref{fig_data}). The substructure
may indeed reflect the blended lines in the 0.7 keV
feature. Investigation of the substructure is in progress with
application of the detailed statistical tests following
\citet{mori05}.

%=============================================================================

\section{Investigation of other proposed models \label{sec_others}} 

Various models have been proposed for the observed spectral
features. We also found several possible solutions involving with
cyclotron lines (case B and C). They span over a large range of
magnetic fields ($B$) and atomic number ($Z$). We ruled out some cases
because self-consistent model spectra with full radiative transfer
solutions exist and they do not reproduce the observed line
parameters. For the other models, we found them implausible because
they lack self-consistency (e.g. ionization balance). Table
\ref{tab_check} summarizes all the models and the reasons for
implausibility.

\subsection{Case A: Pure atomic transition lines} 

\subsubsection{Hydrogen molecular ion model \label{sec_hmolecule}}

At a very high B-field, exotic Hydrogen molecules such as H$_2^+$,
H$_3^{2+}$ and H$_4^{3+}$ become more bound than Hydrogen atoms
\citep{turbiner04}. \citet{turbiner04} attributed the observed
features to photo-ionization and photo-dissociation of Hydrogen
molecular ions at $B=(2-6)\times10^{14}$ G. They predicted several
other spectral features possibly blended in the two broad features.

However, \citet{potekhin04} studied ionization balance of a Hydrogen
atmosphere at $B\uax10^{14}$ G including both atomic and molecular
states. They found that the fraction of Hydrogen molecules is
negligible and atomic hydrogen is predominantly populated at
$T\sim10^6$ K and $B=10^{14}$--$10^{15}$ G. Hydrogen molecules are not
abundant enough to produce observable absorption features in the
spectra. Atomic hydrogen alone cannot produce spectral features at the
observed energies. At $B\uax10^{14}$ G, spectral features from atomic
hydrogen as well as proton cyclotron lines are significantly
suppressed due to vacuum resonance effects \citep{ho03}.

\subsubsection{Helium atmosphere model \label{sec_helium}}

\citet{sanwal02} and \citet{pavlov05} interpret the two features as
atomic transition lines from once-ionized Helium ions at
$B=2\times10^{14}$ G. The transitions considered by \citet{pavlov05}
are the $(00)\rightarrow(01)$ and $\Delta m=1$ transition with an
increment in another quantum number associated with ionic motion for
the 1.4 keV and 0.7 keV feature \citep{pavlov05}. They predicted
several transition lines possibly blended in the 0.7 keV feature.

At such high B-field, the X-ray continuum flux is carried exclusively
by X-mode photons.  Adopting the oscillator strengths of the relevant
transition lines from \citet{pavlov05}, we calculated X-mode
opacities. We found that the $(00)\rightarrow(01)$ transition line is
approximately 2 orders of magnitude smaller than the
$(00)\rightarrow(10)$ transition line at very small $\theta_B$ and
$(00)\rightarrow(01)$ transition is completely suppressed at
$\theta_B\uax10^\circ$. The suppression of tight-loose transitions is
due to the extremely large polarization ellipticity at $B>10^{14}$ G.
It is inconsistent with the comparable line strengths of the two
features (observation (a)). This conclusion is robust regardless of
ionization balance and level population.

The model of \citet{pavlov05} requires that He$^+$ ions largely
populate the atmosphere. However that may not be the case as He atoms
may be abundant and even He molecules may be formed at $B>10^{14}$
G. \citet{turbiner04_2} has shown that He$_2^{3+}$ ion is the most
bound system at $B>10^{14}$ G. Using our atomic code modified for
molecular structure calculations \citep{mori02}, the dissociation
energy of Helium molecules is $\sim$2 keV at $B\sim2\times10^{14}$
G. This is consistent with the results of \citet{lai01}. Since the
thermal energy in the \pulsar atmosphere ($kT\lax200$ eV) is 
$\lax10$\% of the dissociation energy, molecular chains may well be
formed \citep{lai01}. Therefore, transition lines
from He molecules will produce absorption features at different
energies from the observed location.

In addition, vacuum resonance effects may be effective and suppress
spectral features as the vacuum polarization becomes important at
$B>10^{14}$ G.

\subsubsection{Iron atmosphere model  \label{sec_iron}}

\citet{mereghetti02} suggested an Iron atmosphere at $B\sim10^{12}$ G,
although they did not show transition energies corresponding to the
observed features.  We investigated different ionization states of
iron at $B\sim10^{12}$ G, and the only possible combination of
transition lines giving the measured line energy ratio is related to
the $(00)$ and $(10)$ states. However, for all the ionization states,
the binding energy of these states exceeds 10 keV (e.g. $E>1$ MeV for
highly-ionized iron at $B\sim10^{12}$ G). Therefore it is impossible
to match them to the observed features for any reasonable value of
gravitational redshift.  Iron atmosphere should show far more than the
two discrete features in the X-ray band \citep{rajagopal97}.

\subsection{Case B: Pure cyclotron line case}

We find all the pure cyclotron line cases implausible because they are
inconsistent with the observed line strengths and widths. Both
electron and ion cyclotron lines have been previously studied in great
detail and none of the models is consistent with the \xmm data.

\subsubsection{Electron cyclotron line}

\citet{sanwal02} ruled out the electron cyclotron line model because
the 2nd harmonic has significantly smaller line strength than the
fundamental at $B\sim8\times10^{10}$ G. However, \citet{xu05}
argued that electron cyclotron resonant scattering may modify line
strengths resulting in similar strength in the fundamental and the
2nd harmonics.

Radiative transfer models of electron cyclotron lines including the
effects of resonant scattering have been extensively studied for
various physical conditions ranging from NS binaries to gamma-ray
bursts \citep{alexander89,wang93, fenimore88, freeman99}. We
extrapolated these models (especially a semi-analytical model by
\citet{wang93}, hereafter W93) to $B\lax10^{11}$ G, significantly
lower B-fields than the NS binary cases.

We assume a thin isothermal layer composed of fully-ionized plasma
with uniform B-field. We assume that the line-forming plasma is cold
(i.e. $kT \ll \hbar\omega_{eB}$). This simple picture well represents
the proposed electron cyclotron line scenarios either by a layer of
e$^+$e$^-$ or e-p pairs above the NS surface sustained by radiation
pressure \citep{deluca04} or a thin electrosphere on the surface of
bare quark stars \citep{xu05}. The corresponding fit model is model
III and the fitted optical depth ($\tau$) to the \xmm data is
$\sim0.5$ for the both features \citep{mori05}.

\subsubsubsection{Line strength}

Electron cyclotron lines ($\Delta n \ne 0$ transitions) are resonant
lines because de-excitation of electrons from the higher Landau levels
is instantaneous \citep{harding91}. Electrons in $n=1$ states can de-excite
only to $n=0$ states, while electrons in $n=2$ states can either
de-excite directly to $n=0$ states or deexcite to $n=1$ states then to
$n=0$ states (Raman scattering) (W93). The fraction of
Raman-scattered photons from the $n=2$ state is given by $1-B/B_c$
\citep{daugherty77}. At $B\lax10^{11}$ G, electrons in $n=2$ states
will undergo Raman scattering with probability close to unity. This
means that the 2nd harmonic line is absorption-like without suffering
from significant resonant scattering. The fitted optical depth of the
1.4 keV feature immediately gives the electron column density of
$N_e\sim 10^{20}$ cm$^{-2}$ (W93).

On the other hand, the fundamental absorption line is subject to two
different resonant scattering processes: (1) a photon from
$n=1\rightarrow 0$ de-excitation and (2) multiple photons from the
Raman scattering from $n>1$ states. The latter process is called
``spawning'' (W93). At $N_e\sim10^{20}$ cm$^{-2}$, we estimated the
fraction of spawned photons is only $\sim10$\% (W93). The negligible
spawning is due to the fact that the 2nd harmonic line is optically
thin.  On the other hand, the optical depth of the fundamental line is
very large ($\tau\sim 400$) at $N_e\sim10^{20}$ cm$^{-2}$ (W93). A
similar case with such a large optical depth for the fundamental line
is well illustrated in the three right panels in figure 1 of W93. The
figure shows the case of $N_e=1.2\times10^{22}$ cm$^{-2}$ and
$B=1.2\times10^{12}$ G without spawning (accordingly $\tau\sim300$ for
the fundamental line). We note that the model of W93 takes into
account full radiative transfer with resonant scattering.  At any
viewing angle $\theta_B$, the fundamental line appears very deep with
the optical depth far exceeding 1.  This is inconsistent with the
observed optical depth of the 0.7 keV feature ($\tau\sim0.5$).

\subsubsubsection{Line width}

Line widths of the electron cyclotron lines are primarily determined
by the B-field variation over the line-forming region, especially when
the lines are as broad as the \pulsar case. As the cyclotron line
energy is proportional to $B$, $\Delta E/E $ $(\sim \Delta B/B)$ must
be same for the fundamental and 2nd harmonic. This is inconsistent
with the observation (d).

\subsubsection{Ion cyclotron line}

Ion cyclotron lines have been studied in great detail at $B > 10^{14}$
G \citep{zane01, ozel01, ho02}. Ion cyclotron lines at $B>10^{14}$ G
are unlikely to be observable since they are significantly weakened by
vacuum resonance effects in the photosphere \citep{ho02}. Similar to
the electron cyclotron case, the line width argument makes the ion
cyclotron line solutions implausible as well.

\subsection{Case C: mixture of atomic transition and cyclotron line}

All the case C solutions require a large population of H-like ions in
the photosphere because otherwise less ionized states will produce
unobserved features in the X-ray band. The solutions with an electron
cyclotron line are either nitrogen/oxygen at $B\sim2\times10^{11}$ G
or $Z=10$--12 element at $B\sim8\times10^{10}$ G.  However, based on
the ionization balance calculation described in \S\ref{sec_density},
we found that more neutral ions are predominantly populated in the
$(\rho,T)$ range of the \pulsar photosphere.  The solutions with an ion
cyclotron line require either a Helium or Lithium atmosphere at
$B\sim2\times10^{14}$ G. In the former case, He molecular states are
predominantly populated (\S\ref{sec_helium}). We ruled out the latter
case because H-like Li ions are not abundant at all in the photosphere
and also because Lithium abundance is extremely low in the predicted
compositions of supernova ejecta \citep{thielemann96}.

%%%%%%%%%%%%%%%%%%%%%%%%%%%%%%%%%%%%%%%%%%%%%%%%%%%%%%%%%%%%%%%%%%%%%%

\section{Implications of an Oxygen/Neon atmosphere \label{sec_imply}}

The presence of an O/Ne atmosphere makes \pulsar unique while other
INS are likely to have light element atmospheres composed of hydrogen
or helium. There are several possible scenarios for the formation of
an O/Ne atmosphere.  O/Ne may be continuously supplied to the surface
by accretion. As accretion flow from the ISM contains different
elements (e.g., H, He, CNO), and a mixture of atmospheric elements is
expected in the photosphere.  In that case, other elements such as
carbon and nitrogen and possibly helium will produce unobserved
spectral features in the X-ray band. Therefore, we do not consider the
accretion scenario plausible.  We also note that radiative levitation
of O/Ne ions is not effective as the radiative pressure by thermal
flux is insufficient. Therefore, it is natural to assume that \pulsar
has a pure O/Ne atmosphere.

An O/Ne atmosphere could be formed by fallback after a supernova
explosion \citep{herant94}. O/Ne must have survived spallation and
soft-landed on the NS surface \citep{chang04_2}. In the supernova
ejecta, oxygen is a dominantly abundant element over a wide range of
progenitor mass \citep{thielemann96}, so it is feasible that oxygen
would appear on the NS surface rather than neon. Optical observations
establish the existence of oxygen in the vicinity of \pulsar
\citep{ruiz83}. The high galactic latitude of the supernova remnant
PKS 1209-51 indicates an Oxygen-rich environment around the pulsar
\citep{ruiz83}. The fallback scenario rules out a Neon atmosphere and
favors a pure Oxygen atmosphere.

Post-supernova mixing of the ejecta must occur so that oxygen can
diffuse closer to the NS and eventually fall onto the
surface. Large-scale mixing is supported both by observations
\citep{spyromilio94,fassia98} and simulations \citep{hachisu92,
hachisu94, herant94}.  Recent two-dimensional simulations have shown
shortly after a supernova explosion oxygen quickly diffuses toward the
remnant by Rayleigh-Tailor instabilities \citep{kifonidis00,
kifonidis03}. In several particular case simulated by
\citet{kifonidis00} and \citet{kifonidis03}, oxygen and helium are the
most abundant elements in the vicinity of the NS. Note that the degree
of mixing will depend on various factors such as progenitor mass,
delay time of neutrino-driven explosion and mass-cut location
\citep{kifonidis03}.

Along with oxygen, other elements such as hydrogen and helium could
also fall back onto the surface.  \citet{chang04_2} showed a Hydrogen
layer could be completely depleted by diffusive nuclear burning within
7 kyrs (i.e. the age of \pulsar estimated from the surrounding
supernova remnant). Their calculation was performed specifically for
\pulsar by using the NS parameters determined by our present
analysis. Therefore, fallback of hydrogen is not a problem for
\pulsar.  On the other hand, the Helium abundance in the photosphere
must be negligible ($\ll 10^{-19} M_\odot$), otherwise the Helium
layer must be depleted by another mechanism (e.g., pulsar wind) over
the last 7 kyrs.  \citet{chang04_2} estimated that the column density
excavated by pulsar winds is comparable with the amount of hydrogen or
helium in radio pulsars. However, the excavation occurs only at the
polar cap area as ions are ripped off from the surface along open
field lines. Since the polar cap size of \pulsar is small (less than
1km in radius), it is inconsistent with the presence of the absorption
features in all spin phases \citep{deluca04} unless special geometry
is invoked.

Therefore, it is more natural to assume that a pure Oxygen layer was
formed soon after the supernova explosion.  Presence of oxygen on the
surface of \pulsar puts stringent conditions on the post-supernova
environment and the fallback mechanism. All of the following processes
(i.e. formation of NS, mixing of an Oxygen shell toward the NS,
negligible Helium contamination and soft-landing of Oxygen nuclei
without spallation) must occur after the supernova explosion.

%==============================================================================

\section{Summary \label{sec_last}} 

\begin{itemize}

\item An O/Ne atmosphere at $B\sim10^{12}$ G is the only consistent
solution with the observed line parameters.

\item Other proposed models are ruled out either because they are
inconsistent with the observed line parameters or because they lack
self-consistency (e.g. ionization balance).

\item The magnetic field strength and gravitational redshift are
uniquely determined: $B_{12}\simeq0.5$--$1.0, z\simeq0$--$0.4$ for
oxygen, and $B_{12}\simeq 1.0$--2.0, $z\simeq0.4$--0.8 for neon.

\item The two broad features are due to bound-bound transitions from
highly-ionized O/Ne. Bound-free and free-free absorption are not
important.

\item Highly-ionized (H-, He- and Li-like) O/Ne are largely populated
in the photosphere.

\item The observed features are broad due to motional Stark effects
and magnetic field variation over the surface.

\item Several photo-absorption lines from different O/Ne ions are
blended in the two features. In our model, the 0.7 keV feature has
larger line width than the 1.4 keV feature and may show substructure in
the \xmm/EPIC spectra because it contains more blended absorption
lines.

\item Mixed atmosphere composition is allowed only with hydrogen and
only if O/Ne is continuously supplied by accretion. However, the
admixture atmosphere case is implausible because accretion flow from
the ISM will contain other mid-Z elements (e.g. carbon and nitrogen).

\item Fallback of supernova ejecta is the most plausible scenario for
the formation of an O/Ne atmosphere. A Neon atmosphere is ruled out
because the Neon abundance is much smaller than the Oxygen abundance
in the supernova ejecta. Large-scale mixing of supernova ejecta and
soft-landing fallback of oxygen with negligible Helium contamination
must have occurred after the explosion.

\end{itemize} 

In the present analysis, we have constrained the surface composition
of \pulsar using the CCD spectroscopy aboard \xmm telescope. Resolving
blended lines by high resolution grating spectroscopy will lead to
highly accurate measurements of magnetic field strength and
gravitational redshift to better than 10\% accuracy. Phase-resolved
spectroscopy will possibly decouple magnetic field effects from
gravitational effects because the magnetic field varies with spin
phase, while gravitational redshift does not. Radiative transfer
using the LTE opacities is in process for calculating self-consistent
temperature profiles and emergent spectra \citep{mori06_2}. Our spectral
models will be fitted to the \xmm data and determine NS parameters.

\begin{acknowledgements} 

We thank W. Ho, D. Lai, G. Pavlov, A. Potekhin, C. Thompson, A. Turbiner and
M.H. van Kerkwijk for discussions and P. Chang for calculating diffuse
nuclear burning rate for \pulsar.

\end{acknowledgements}

%\bibliography{/cita/h/home-3/kaya/apj/ns}
%\bibliographystyle{/cita/h/home-3/kaya/apj/apj.bst}

%=============================================================================
% Figures 
%=============================================================================

%\clearpage

%=============================================================================
% Tables 
%=============================================================================

\begin{deluxetable}{cccccccc}
\tablewidth{0pt}
\tablecaption{Unredshifted line energies [keV] for H-, He- and Li-like oxygen in the range of $B_{12}=0.5$--1.  \label{tab_line}} 
\tablehead{
\colhead{Transition} & \colhead{$(00)\rightarrow (01)$} & \colhead{$(00)\rightarrow(10)$ \tablenotemark{a}} & \colhead{$(10)\rightarrow (11)$} & \colhead{$(10)\rightarrow (02)$} & \colhead{$(20)\rightarrow (21)$ \tablenotemark{b}} &   \colhead{$(20)\rightarrow (12)$ \tablenotemark{c}}} 
\startdata
Which feature? & 1.4 keV & 0.7 keV & 0.7 keV & 0.7 keV & 0.7 keV & 0.7 keV \\ 
\hline 
Polarization mode ($\alpha$) & 0 & +1 & 0 & -1 & 0  & -1 \\ 
\hline 
\ion{O}{8}	& 1.51--1.98 	&  0.81--0.95 & 0.79--1.12 &  0.96--1.32 &  0.58--0.84 & 0.74--1.04 \\
\ion{O}{7}	& 1.46--1.91 	&  0.77--0.91  & 0.74--1.03  & 0.87--1.19 &  0.52--0.76 &  0.65--0.92\\
\ion{O}{6}	& 1.42--1.84	&  0.76--0.89  & 0.71--0.99 &  0.82--1.12 &   0.49--0.71 & 0.60--0.83    
\enddata
%\tablecomments{The Neon case requires slightly higher B-field ($B_{12}\simeq1$--2) and gravitational redshift ($z\simeq0.4$--0.8).}
\tablenotetext{a}{$(00)(20)\rightarrow(10)(20)$ for He-like ions. $(00)(10)(30)\rightarrow(00)(20)(30)$ for Li-like ions. }
\tablenotetext{b}{$(00)(20)\rightarrow(00)(21)$ for He-like ions. $(00)(20)(30)\rightarrow(10)(20)(30)$ for Li-like ions.}
\tablenotetext{c}{$(00)(20)\rightarrow(00)(12)$ for He-like ions. $(00)(20)(30)\rightarrow(00)(12)(30)$ for Li-like ions.}
\end{deluxetable}

\begin{deluxetable}{ccccccc}
\tablewidth{0pt}
\tablecaption{Check list of  all the models considered. \label{tab_check}}
\tablehead{\colhead{Case} & \colhead{Element} & \colhead{Ionization states} & \colhead{B [G] \tablenotemark{a}} & \colhead{0.7 keV} & \colhead{1.4 keV} & Why implausible?}
\startdata
A       & O or Ne & H-, He-, Li-like & $10^{12}$ & L & L & OK\\
A       & H & H$_3^{2+}$, H$_4^{3+}$ & $4\times10^{14}$ &  &  & IB \& VR \\  
A       & He & H-like & $2\times10^{14}$ & L & L & LS, IB \& VR \\ 
A       & Fe &  unknown      & $10^{12}$ & ? & ? & Too many features \\    
B       & -- & Fully-ionized  & $8\times10^{10}$ & EC & EC & LS \& LW     \\
B       & -- & Fully-ionized  & $6\times10^{14}$ & IC& IC & LS \& LW  \\ 
C       & N or O & H-like & $10^{11}$ & L & EC  & IB        \\
C       & Ne, Na or Mg & H-like & $8\times10^{10}$ & EC & L & IB           \\
C       & He or Li & H-like & $2\times10^{14}$  & L & IC &  IB \& VR    
\enddata
\tablecomments{L - photo-absorption line, EC - electron cyclotron line and IC - ion cyclotron line, LS - Line strength, LW - Line width, IB - Ionization balance and VR - Vacuum resonance effect}  
\tablenotetext{a}{B-field values are approximate due to uncertainty associated with gravitational redshift.  } 
\end{deluxetable}

\begin{deluxetable}{cccc}
\tablewidth{0pt}
\tablecaption{Estimated line widths ($\Delta E/E$ [\%]) by various line broadening mechanisms. \label{tab_broad}} 
\tablehead{\colhead{Mechanism} & \colhead{0.7 keV} & \colhead{1.4 keV} & \colhead{Relevant parameters}}
\startdata
Doppler broadening & $<0.1$ & $<0.1$ & $T$, spin period \\ 
Pressure broadening 	& $< 1$ & $<1$ & $\rho, T$	\\
Motional Stark broadening    &  $\lax$5    & $\lax$1 & $B$, $T$	\\
Different ionization states & $\lax20$   &  $\lax20$ & $B$,  $T$\\
B-field variation\tablenotemark{a}  &  $\sim 7$--14 \tablenotemark{b}  & $\sim 12$ & $B$ 	\\
\enddata 
\tablenotetext{a}{Line widths correspond to B-field variation by 30\% (from $B_{12}=0.75$ to $B_{12}=1.0$).}
\tablenotetext{b}{7\% and 14\% line width correspond to $(00)\rightarrow(10)$ transition of H-like ions and $(00)(10)\rightarrow(00)(11)$ transition of He-like ions respectively. } 
\tablecomments{The measured line widths are $\sim 40$ \% and $\sim 15$ \% for the 0.7 and 1.4 keV feature. }
\end{deluxetable}

\end{document}